\documentclass[10pt,final,journal,letterpaper,twoside,twocolumn]{IEEEtran}

\usepackage{amsmath,amssymb,amsfonts}
\usepackage{cite}
\usepackage{color}
\usepackage{graphicx,graphics}
\usepackage{subfigure,dblfloatfix}

\newtheorem{theorem}{Theorem}

\newtheorem{corollary}{Corollary}

\newtheorem{lemma}{Lemma}
\newtheorem{remark}{Remark}

\newcommand{\PP}{\Pr}
\newcommand{\Prob}[1]{\PP [#1]}

\providecommand{\abs}[1]{\lvert#1\rvert}

\renewcommand{\t}[1]{\ensuremath{\Theta (#1)}}

\newcommand{\set}[1]{\mathcal{#1}}
\def\func{\operatorname}
\newcounter{MYtempeqncnt}
\def\myfigwidth{10.8cm}
\renewcommand{\IEEEQED}{\IEEEQEDopen}

\begin{document}

\title{Opportunistic Relaying in Wireless Networks}
\author{Shengshan~Cui,~\IEEEmembership{Member,~IEEE,}
Alexander~M.~Haimovich,~\IEEEmembership{Senior~Member,~IEEE,}\\
Oren Somekh,~\IEEEmembership{Member,~IEEE,}
and H.~Vincent~Poor,~\IEEEmembership{Fellow, IEEE}%
\thanks{%
Manuscript submitted December 6, 2007; revised January 6, 2009 and May 31, 2009.}%
\thanks{%
The work of S.~Cui and A.~M.~Haimovich was supported in part by the
National Science Foundation under Grant CNS-0626611. The work of
O.~Somekh was supported in part by a Marie Curie Outgoing
International Fellowship within the 6th European Community Framework
Programme. The work of H.~V.~Poor was supported in part by National
Science Foundation under Grants ANI-0338807 and CNS-0625637. The
material in this paper was presented in part at the 45th Annual
Allerton Conference on Communications, Control and Computing,
Monticello, IL, USA, September 2007, and the IEEE International Symposium
on Information Theory, Toronto, ON, Canada,
July 2008.}%
\thanks{%
S.~Cui was with the Department of Electrical and Computer
Engineering, New Jersey Institute of Technology, Newark, NJ 07102
USA. He is now with Qualcomm Inc., San Diego, CA 92121 USA (e-mail: scui@qualcomm.com).}%
\thanks{%
A.~M.~Haimovich is with the Department of Electrical and Computer
Engineering, New Jersey Institute of Technology, Newark, NJ 07102
USA (e-mail: alexander.m.haimovich@njit.edu).}%
\thanks{%
O.~Somekh was with the Department of Electrical
Engineering, Princeton University, Princeton, NJ 08544 USA. He is now with
with Department of Electrical Engineering, Technion--Israel Institute of
Technology, Technion City, Haifa 32000, Israel
(e-mail: orens@princeton.edu).}%
\thanks{%
H.~V.~Poor is with the Department of Electrical
Engineering, Princeton University, Princeton, NJ 08544 USA (e-mail:
poor@princeton.edu).}
}%


\maketitle

\newpage
\begin{abstract} \boldmath
Relay networks having $n$ source-to-destination pairs and $m$
half-duplex relays, all operating in the same frequency band and in the
presence of block fading, are analyzed. This setup has attracted
significant attention, and several relaying protocols have been
reported in the literature. However, most of the proposed solutions
require either centrally coordinated scheduling or detailed channel
state information (CSI) at the transmitter side. Here, an
opportunistic relaying scheme is proposed that alleviates these
limitations, without sacrificing the system throughput scaling in the
regime of large $n$. The scheme entails a two-hop communication
protocol, in which sources communicate with destinations only
through half-duplex relays. All nodes operate in a completely
distributed fashion, with no cooperation. The key idea is to
schedule at each hop only a subset of nodes that can benefit from
\emph{multiuser diversity}. To select the source and destination
nodes for each hop, CSI is required at receivers (relays for
the first hop, and destination nodes for the second hop), and an
index-valued CSI feedback at the transmitters. For the case when
$n$ is large and $m$ is fixed, it is shown that the proposed scheme
achieves a system throughput of $m/2$ bits/s/Hz. In contrast, the
information-theoretic upper bound of $(m/2)\log \log n$ bits/s/Hz is
achievable only with more demanding CSI assumptions and cooperation
between the relays. Furthermore, it is shown that, under the
condition that the product of block duration and system bandwidth
scales faster than $\log n \log\log n$, the achievable throughput of
the proposed scheme scales as $\Theta \left( {\log n}\right)$.
Notably, this is proven to be the optimal throughput scaling even if
centralized scheduling is allowed, thus proving the optimality of
the proposed scheme in the scaling law sense. Simulation results
indicate a rather fast convergence to the asymptotic limits with the
system's size, demonstrating the practical importance of the scaling
results.
\end{abstract}


\section{Introduction}

\label{sec:introduction}

\IEEEPARstart{T}{he demand} for ever larger and more efficient wireless
communication networks necessitates new network architectures, such
as \emph{ad hoc} networks and relay networks. As such, there has
been significant activity in the past decade toward understanding
the fundamental system throughput limits of such architectures and
developing communication schemes that seek to approach these limits.

Among other notable recent results on the throughput scaling of
wireless networks, Gowaikar \emph{et al.}\ \cite{GHH:06} proposed a
new wireless ad hoc network model, whereby the strengths of the
connections between nodes are drawn independently from a common
distribution, and analyzed the system throughput under
various fading distributions. Such a model is appropriate for
environments with rich scattering but small \emph{physical} size, so
that the connections are governed by random fading instead of
deterministic path loss attenuations. When the random channel
strengths follow a Rayleigh fading model, the system throughput
scales as $\log n$. This result is achievable through a multihop
scheme that requires central coordination of the routing of nodes.
Moreover, full knowledge of the channel state information (CSI) of
the entire network is needed to enable the central coordination.

Along with the work on multihop schemes, such as \cite{GHH:06} and
\cite{GK:00}, there is another line of work characterizing the
system throughput for wireless networks operating with two-hop
relaying. The \emph{listen-and-transmit} protocol, studied by Dana
and Hassibi \cite{DH:06} from the power-efficiency perspective,
turns out to have interesting properties from the system
throughput standpoint as well. This is in fact a two-hop
\emph{amplify-and-forward} scheme, where relays are allowed to
adjust the phase and amplitude of the received signals. A throughput
of $\Theta(n)$ bits/s/Hz is achieved by allowing $n$
source-to-destination (S--D) pairs to communicate, while
$m=\Theta(n^{2})$ nodes in the network act as relays. It is assumed
that each relay node has full knowledge of its local channels
(backward channels from all source nodes, and forward channels to
all destination nodes), so that the relays can perform
\emph{distributed beamforming}. Morgenshtern and B{\"{o}}lcskei
worked in \cite{MB:07} with a similar distributed beamforming setup,
and their results reveal trade-offs between the level of available
channel state information and the system throughput. In particular,
utilizing a scheme with relays partitioned into groups, and where
relays in each group have CSI knowledge of only one backward and one
forward channel, the number of relays required to support a $\Theta
\left( n\right) $ throughput is $m=\Theta( n^{3})$. In other words,
with lower level CSI, the number of required relays increases from
$\Theta(n^{2})$ to $\Theta(n^{3})$. An equivalent point of view is
to state the throughput in terms of the total number of transmitting
nodes in the system, $p=n+m$. Then the system throughput is $\Theta
\left( p^{1/3}\right)$, when the relays in each group know the
channel for only one source-destination pair. When relays know the
channels for all source and destination nodes, the throughput scales
as $\Theta \left( p^{1/2}\right) $.

Although these works have made great strides toward understanding
wireless ad hoc network capacity, implementations of the schemes
require either central coordination among nodes \cite{GK:00,GHH:06}
or some level of CSI (channel amplitude and/or phase) at the
transmitter side \cite{DH:06,MB:07}. The centralized coordination
between wireless relays does not come for free, since the overhead
to set up the cooperation may drastically reduce the useful
throughput \cite{OLT:07}.\footnote{However, in throughput scaling
law studies, see, e.g., \cite{GK:00,GHH:06} and \cite{OLT:07}, among many
others, the overhead needed to set up cooperation is usually not
explicitly accounted for.} Likewise, in a large system, obtaining
this level of CSI, especially at the transmitter side, may not be
feasible. This paper addresses the need to alleviate these
limitations by proposing an opportunistic relaying scheme that works
in a completely decentralized fashion and imposes less stringent CSI
requirements.

\subsection{Main Contributions and Related Work}

The main contributions of this work can be summarized as follows.

\begin{itemize}
\item A two-hop opportunistic relaying scheme for operating over fading
channels is proposed and analyzed. The scheme's salient features
are:

\begin{itemize}
\item[---] It operates in a decentralized fashion. No cooperation
among nodes is assumed or required.

\item[---] Only modest CSI requirements are imposed. At each hop, each receiver
is assumed to have knowledge of its local incoming channel
realizations, while transmitters have access to only index-valued
CSI via low-rate feedback from the receivers.
\end{itemize}

\item The throughput of the proposed scheme is characterized by:

\begin{itemize}
\item[---] It is shown that, in the regime of a large number of
nodes $n$ and fixed number of relays $m$, the proposed scheme
achieves a system throughput of $m/2$ bits/s/Hz. This can be
contrasted with the information-theoretic upper bound $(m/2)\log
\log n$ on the scaling of the throughput, achievable only with full
cooperation among the relays and full CSI (backward and forward) at
the relays. These results reveal an interesting feature of multiuser
diversity: whereas full cooperation between relays can readily
form parallel channels, and multiuser diversity can boost the
throughput of each channel by a factor of $\log \log n$,
when cooperation is not possible, multiuser diversity succeeds in
restoring the parallel channels, but must forsake the multiuser
diversity factor $\log\log n$.

\item[---] We show that $m$ can grow (as a function of $n$) as fast as
$\Theta(\log n)$, while still guaranteeing the linear throughput
scaling in $m$. The linearity breaks down if $m$ grows faster than
$\Theta (\log n)$. Furthermore, when the product of the block
duration and the system bandwidth scales faster than $\log n\log\log
n$, the overhead due to feedback is negligible, and therefore, the
achievable throughput scaling of the proposed opportunistic relaying
scheme is given by $\Theta (\log n)$.
\end{itemize}

\item It is proven that, under the assumption of independent
encoding (i.e., no cooperative encoding) at the transmitters and
independent decoding (i.e., no cooperative decoding) at the
receivers, the system throughput is upper-bounded by the order of
$\log n$, even if centralized scheduling is allowed. This result is
of interest in its own right, since it quantifies the system
throughput of wireless ad hoc networks under the scenario where
neither transmitters nor the receivers can cooperate in avoiding
and/or canceling interference. Thus, the network is
interference-limited, unlike other works in which global CSI is
assumed (and thus either cooperative encoding or decoding is
possible), leading to a linear throughput scaling. The throughput
scaling results under our pessimistic, yet more realistic, scenario,
improve the understanding of throughput scaling of wireless
ad hoc networks.

\item The proposed scheme is order-optimal in achieving the $\Theta(\log n)$ throughput scaling.
This suggests that, as far as throughput scaling is concerned,
operating the network in a decentralized fashion, with local CSI at
the receivers and low-rate feedback, incurs no loss.

\item Simulation results show that the asymptotic conclusions
developed in this work settle rapidly. Hence, the above scaling laws
provide rule-of-thumb guidance for the design of practical wireless systems.
\end{itemize}


The key idea behind the proposed scheme is to schedule at each hop
only the subset of nodes that can benefit from \emph{multiuser
diversity}. The concept of multiuser diversity was originally
studied in the context of cellular systems
\cite{KH:95,Tse:97,VTL:02}. It is known that the capacity of
single-cell system is maximized by allowing only the user with the
best channel to transmit at any given time. The concept is by now
well understood in the context of infrastructure wireless networks,
and has been adopted in 3G cellular systems and other emerging
wireless standards \cite{Bender_etal:00}. However, to the best of
our knowledge, it has received less attention for wireless ad hoc
networks, with some exceptions such as \cite{BKRL:06}, in which the
potential of opportunistic relaying is reported in a setup with one S--D pair and
multiple relay nodes, and the focus is on diversity-multiplexing trade-off analysis \cite{ZT:03}.
In this work, we highlight another aspect of
multiuser diversity:\ its application to simplify network operations
and its effect on throughput scaling. The opportunistic scheme
proposed here is in the spirit of \cite{GT:02}, where
distance-dependent, random channel gains were exploited in
scheduling.

In this work, we restrict ourselves to those assumptions that are
implementable with the state-of-the-art technologies. Specifically,
we focus on the assumptions of perfect CSI at the receivers and
partial CSI at the transmitters via low-rate feedback. With these
less idealistic CSI assumptions, it is envisioned that independent
encoding at the transmitters and independent decoding at the
receivers are employed. To the best of the authors' knowledge, there
are no analogous results in the literature that consider the same
scenario. It was recently shown, however, that under more optimistic
assumptions on CSI in the network, linear throughput can be achieved
by either joint encoding at the transmitters \cite{CJ:08} or
hierarchical cooperation with joint decoding at the receivers
\cite{OLT:07}.

\subsection{Organization of the Paper}

The rest of the paper is organized as follows. The system model and
the proposed two-phase relay protocol are introduced in
Section~\ref{sec:sys_model}. Section~\ref{sec:fix_m_large_n}
characterizes the system throughput in the regime of large $n$
and fixed $m$. The throughput scaling of the proposed scheme is
evaluated in Section~\ref{sec:large_m_large_n} by explicitly taking
the feedback overhead into account. Also in
Section~\ref{sec:large_m_large_n}, the throughput upper bound is developed, valid
even when centralized scheduling is allowed.
Section~\ref{sec:simulation} presents numerical performance results.
Section~\ref{sec:discussions} briefly discusses the impact of relay
cooperation on system throughput and the delay consideration. Finally,
Section~\ref{sec:conclusion} concludes the paper. Technical
details and proofs are placed in the appendices.

\emph{Notation}: The symbol $\abs{\set{X}}$ denotes the cardinality
of the set $\mathcal{X}$, and unless specified otherwise,
$\log(\cdot)$ indicates the natural logarithm. We write $X\sim
\func{Exp}(1)$ to indicate that the random variable $X$ follows the
standard exponential distribution with probability density function
(pdf) given by $f_{X}(x)=\exp (-x),\,x>0$. The indicator function is
denoted by $\mathbf{1} (\cdot )$, and we use \textquotedblleft
$\chi^{2}(2p)$\textquotedblright\ to denote a chi-square random
variable with $2p$ degrees of freedom. For two functions $f(n)$ and
$g(n)$, $f(n)=O(g(n))$ means that $\lim_{n\to \infty}
\abs{f(n)/g(n)} <\infty$, and $f(n)=\Omega(g(n))$ means that
$g(n)=O(f(n))$. We write $f(n)=o(g(n))$ to denote $\lim_{n\to
\infty} \abs{f(n)/g(n)} = 0$, and $f(n)=\Theta(g(n))$ to denote
$f(n)=O(g(n))$ and $g(n)=O(f(n))$.

\section{System Model}

\label{sec:sys_model}

\begin{figure*}[htb!]
\centerline{ \subfigure[]{\includegraphics{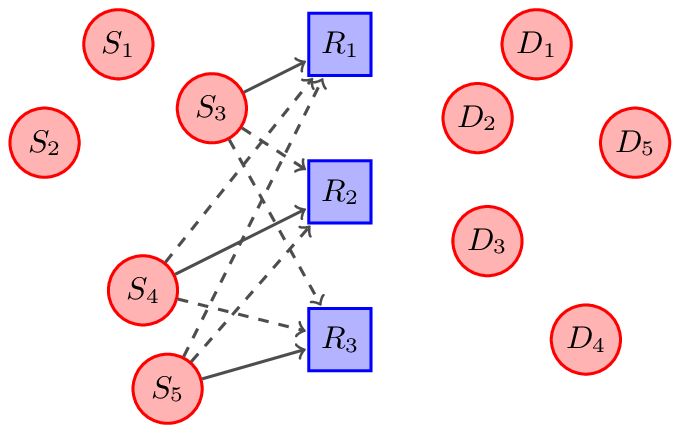}}
\hfil \subfigure[]{\includegraphics{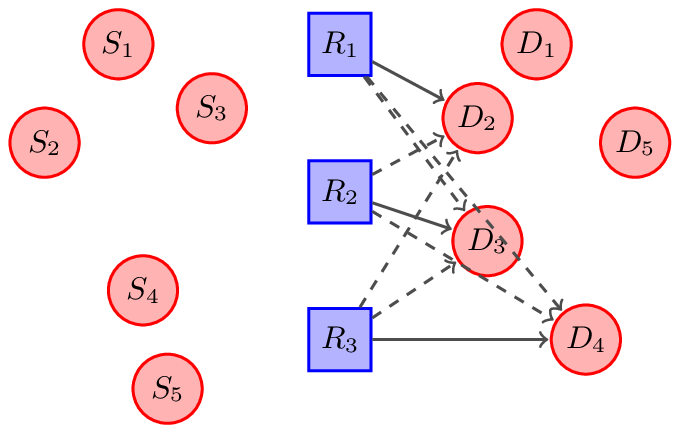}}}
\caption{\label{fig:sys_model}A two-hop network with $n=5$ S--D
pairs and $m=3$ relay nodes (denoted by the blue disks). (a) In the
first hop, source nodes $\{3,4,5\}$ transmit to the relays. (b) In
the second hop, the relays transmit to the destination nodes
$\{2,3,4\}$. Solid lines indicate scheduled links, while dashed
lines indicate interfering links.}
\end{figure*}

Consider a wireless network with $n$ S--D pairs and $m$ relay nodes,
all operating in the same frequency band of width $W$ Hz, in the
presence of fading. Ad hoc nodes that generate data traffic are
referred to as source nodes; nodes that receive data traffic are
referred to as destination nodes. Relay nodes have no intrinsic
traffic demands. We consider a two-hop, decode-and-forward
communication protocol, in which sources can communicate with their
destinations only through half-duplex relays. In the first hop of
the protocol, a subset of sources is scheduled for transmission to
relays. The relays decode and buffer the received packets. During
the second hop of the protocol, the relays forward packets to a
subset of destinations (not necessarily the set of destinations
associated with the source set in the first hop). These two phases
are interleaved: the first hop is
run in the even-indexed time-slots, and the second hop is run in the
odd-indexed time-slots. An
example of the two-hop relay protocol is depicted in
Fig.~\ref{fig:sys_model}.

We first describe the channel model. It is assumed that the wireless
network has independent and identically distributed (i.i.d.)
Rayleigh connections $h_{i,r}$ from source nodes $i$, $1\leq i\leq
n$, to relay nodes $r$, $1\leq r\leq m$. Thus, the channel gains
follow an exponential distribution, i.e.,
$\gamma_{i,r}=\abs{h_{i,r}}^{2}\sim \func{Exp}(1)$. Likewise, we
assume that the channel gains $\xi_{r,j}$ from relays $r$, $1\leq
r\leq m$, to destination nodes $j$, $1\leq j\leq n$, be i.i.d.
$\func{Exp}(1)$, and that channel gains $\gamma_{i,r}$ and $\xi
_{r,j}$ are independent for all $i$, $r$, and $j$. This model is
appropriate for dense networks in a rich scattering environment,
where the distance between transmitters and receivers has only a
marginal effect on attenuation, and the channel attenuation is
dominated by the small-scale fading due to multipath.
Quasi-static fading is assumed, with channel gains fixed during the
transmission of each hop, which is assumed to have a duration of $T$
seconds, and taking on independent values at different transmission
times. In practice, $T$ can be as large as the coherence time of the
channel allows. Regarding CSI, we assume that at each hop, each
receiver has perfect CSI knowledge, while the
transmitters have access only to an \emph{index value} via receiver
feedback used to indicate a source chosen for transmission. This CSI
assumption is reasonable in practice as most wireless access network
standards incorporate some form of pilot signals, and the type of
feedback specified has low overhead.

We now describe the scheduling at each hop. We start with the first
hop (Phase $1$). All relays operate independently. Thus, without
loss of generality, let us focus on any specific relay, say $r$. By
assumption, relay $r$ has the knowledge of $\gamma_{i,r},
i=1,\ldots,n$, and it will schedule the transmission of the strongest
source node, say $i_r=\arg \max_{i}\gamma_{i,r}$, by feeding back
the index $i_r$ at the beginning of the block. The overhead of this
phase of the protocol is a single integer per relay node. Suppose
the scheduled nodes constitute a set $\mathcal{K}\subset \{1,\ldots
,n\}$; then since there are $m$ relays, up to $m$ source nodes can
be scheduled in this fashion, i.e., $\abs{\set{K}}\leq m$. It is
noted that it is possible for multiple relays to schedule
the same source. In such cases, $\abs{\set{K}}<m$. The scheduled
source nodes transmit simultaneously with constant power $P$ and
fixed transmission rate of $1$ bit/s/Hz.\footnote{Generalizing to
higher transmission rates is straightforward, but it encumbers
notation without adding insight. See discussions in
Remark~\ref{rmk:1bps} for the motivation of choosing the rate of $1$
bit/s/Hz.} Each relay sees a superposition of all the transmitting
signals, i.e.,
\begin{equation*}
y_r=\sqrt{P}\,h_{i_{r},r}\,x_{i_r} + \sum_{\substack{t\in
\mathcal{K}\\t\neq i_r}}\sqrt{P}\,h_{t,r}\,x_{t}+n_{r},\quad
r=1,\ldots, m
\end{equation*}
where $x_{i}$ denotes the transmitted signal of source $i$, $n_r$
denotes the additive noise at relay $r$. In this paper, we assume
$x_{i}$'s are letters from codewords of a Gaussian
capacity-achieving codebook satisfying $\mathbb{E}[\abs{x_i}^2]=1$.
We further assume $n_r$'s are i.i.d. complex Gaussian with zero mean
and unit variance $\mathcal{CN}(0,1)$ and are independent of the
fading channels. Since the transmission rate is $1$ bit/s/Hz,
communication can be supported in an information-theoretic sense if
the corresponding signal-to-interference-plus-noise ratio (SINR) is
greater or equal to one, i.e.,
\begin{equation}
\mathsf{SINR}_{i_{r},r}^{\mathrm{P1}}=\frac{\gamma_{i_{r},r}}{1/\rho
+ \sum_{\substack{ t\in \mathcal{K} \\ t\neq
i_{r}}}\gamma_{t,r}}\geq 1, \quad r=1,\ldots, m \label{eq:sinr_ul}
\end{equation}%
where $\rho=P$ is the average signal-to-noise ratio (SNR) of the
source--relay link.

The scheduling at the second hop (Phase $2$) works as follows. All
relay nodes transmit simultaneously with fixed power $P_R$. Assume
that the additive noise at the destination nodes are i.i.d.
$\mathcal{CN}(0,1)$ and are independent of the fading processes
$\{\xi_{r,j}\}$, and assume that independent messages are sent
across relay nodes (which is the case in the proposed scheduling),
destination node $j$ can compute $m$ SINRs by assuming that relay
$r$ is the desired sender and the other relays are interference as
follows:
\begin{equation}
\mathsf{SINR}_{r,j}^{\mathrm{P2}}=\frac{\xi _{r,j}}{1/\rho _{R}+\sum
_{\substack{ 1\leq \ell \leq m \\ \ell \neq r}}\xi _{\ell ,j}},
\quad r=1,\ldots, m \label{eq:sinr_dl}
\end{equation}%
where $\rho_{R}=P_R$ denotes the average SNR of a relay--destination
link. If the destination node $j$ captures one good SINR, say,
$\mathsf{SINR}_{k,j}^{\mathrm{P2}}\geq 1$ for some $k$, it instructs
relay $k$ to send data by feeding back the relay index $k$ at the
beginning of the block. Otherwise, the node $j$ does not provide
feedback. It follows that the overhead of the second hop is also
{\em at most} an index value per destination node. When scheduled by
a feedback message, relay $k$ relays the data to the destination
node at rate $1$ bit/s/Hz. In case a relay receives multiple
feedback messages, it randomly chooses one destination for
transmission. It is noted that in steady-state operation of the
system, the relays have the ability to buffer the data received from
source nodes, such that it is available when the opportunity arises
to transmit it to the intended destination nodes over the second hop
of the protocol. This ensures that relays always have packets
destined to the nodes that are scheduled. In addition, due to the
opportunistic nature of scheduling, the received packets at the
destinations are possibly out of order, and therefore each
destination is assumed to be able to buffer them before decoding.
\smallskip
\begin{remark}
\label{rmk:P1_vs_P2} It is noteworthy to draw a comparison between
Phase $1$ and Phase $2$. From the relays' perspective, both hops of
the communication protocol rely on scheduling a subset of
\textquotedblleft good\textquotedblright\ source/destination nodes
for transmission. However, these two phases of the protocol differ
in one key aspect: transmission over the second hop can be
guaranteed to be successful since receivers (the destinations) have
access to the SINRs, but this is not the case for the first hop.
This is because in the first hop, each receiver (relay) selects a
source node without knowledge of what the other relays select. As a
consequence, each relay has no access to the interference stemming
from all other concurrent transmitting sources, and therefore has no
\emph{a priori} knowledge of its own SINR. For example, in
\eqref{eq:sinr_ul} relay $r$ knows the desired link strength
$\gamma_{i_r,r}$, but it does not know $\mathcal{K}$ and the
corresponding interference term $\sum_{t\in \mathcal{K},t\neq
i_{r}}\gamma_{t,r}$. For the second hop, the senders (now the
relays) are known \emph{a priori}, and therefore the destination
nodes have direct access to SINRs. This implies that once the
destination node captures an $\mathsf{SINR}^{\mathrm{P2}}\geq 1$,
and accordingly requests a transmission, this transmission will be
successful at a data rate of $1$ bit/s/Hz. This key difference
between the two phases is mirrored in the analysis in
Section~\ref{sec:fix_m_large_n}.
\end{remark}
\smallskip
\begin{remark}
In both hops, {\em independent encoding} at the transmitters and
{\em independent decoding} at the receivers are employed. By
independent encoding, it is meant that the transmitters encode their
message independently. This is a consequence of the fact that the
transmitters have access to only partial CSI. Similarly, by
independent decoding is meant that receivers decode their message
independently by treating interference as noise. It is worth
pointing out that, with the assumption of CSI at the receivers,
techniques like interference cancelation are possible at the
receivers. However, as elaborated in Remark~\ref{rmk:1bps}, they are
not interesting in our setup and thus are not considered here.
\end{remark}

\section{Throughput: Large $n$ and Fixed $m$}

\label{sec:fix_m_large_n}

Motivated by the observation that as communication devices (source
and destination nodes in our system) become more and more pervasive,
the number of infrastructure nodes (relays) is not likely to keep
pace, the throughput analysis in this paper pays special attention
to a regime in which the number of source and destination nodes,
$n$, is large, while the number of relay nodes, $m$, is relatively
small. We show that both Phase $1$ and Phase $2$ achieve
average throughput (by averaging over random channel gains) of $m$
bits/s/Hz, yielding a $m/2$ bits/s/Hz throughput for the complete
two-hop scheme. We also show that for \emph{any} two-hop protocol,
the throughput is upper-bounded by $\frac{m}{2}\log \log n$
bits/s/Hz. This information-theoretic upper bound holds even if we
allow full cooperation between relays \emph{and} assume full CSI is
available at the relays. Thus, the proposed scheme, with much
simplified assumptions of decentralized relay operations and CSI at
the receiver, succeeds in maintaining the linearity of the
throughput in the number of relay nodes.

\subsection{Phase $1$: Source Nodes to Relays}

\label{sec:Pone_large_n_fix_m}

In Phase $1$, $m$ relays operate independently and each schedules
one source node for transmission. Hence, the total number of
scheduled source nodes can be any integer between $1$ and $m$, i.e.,
$\abs{\set{K}}\leq m$. In cases when $\abs{\set{K}}<m$, multiple
relays schedule the same source node, and the analysis of the
probability of successful transmission should consider explicitly
those links with multiple receivers. Due to the multiplicity of
possible combinations, the exact characterization of the average
throughput of Phase $1$, $R_{1}$, is mathematically involved.
Fortunately, in order to show the achievability of $m$ successful
concurrent transmissions, it suffices to lower-bound $R_{1}$ by
considering only cases in which the $m$ scheduled source nodes are
distinct (thereby discarding the contributions to the throughput of
the other combinations).

By symmetry, each source node has a probability of $1/n$ to be the
best node with respect to a relay. Thus, the probability that the
scheduled users are distinct, i.e., no source node is scheduled by
more than one relay, is given by $\Pr [N_{m}]=n(n-1)\cdots
(n-m+1)/n^{m}$, where $N_{m}$ denotes the event $\{$$m$ distinct
source nodes are scheduled$\}$. Now, a lower bound on $R_{1}$ is
\begin{align}
R_{1}& \geq \Pr \bigl[N_{m}\bigr]\sum_{r=1}^{m}\Pr \bigl[\mathsf{SINR}_{i_{r},\,r}\geq 1\bigr]\cdot 1 \notag\\
&=m \, \Pr\bigl[N_m\bigr] \Pr \bigl[ \mathsf{SINR}^{\mathrm{P1}}\geq
1\bigr], \label{eq:lower_bound_R_1}
\end{align}%
where, for notational brevity and by the i.i.d. channel model, we
drop the source node and relay indices in the last equation and use
$\mathsf{SINR}^{\mathrm{P1}}$ to denote the SINRs at all relay
nodes.

Now, we focus on the $\Pr \bigl[ \mathsf{SINR}^{\mathrm{P1}}\geq
1\bigr]$ term. Again, for notational convenience, for a realization
of $n$ i.i.d. random variables $X_{1},\ldots,X_{n}$, we introduce
$X:=\max\{X_{1},\ldots,X_{n}\}$ and $Y:=\sum_{i\in \mathcal{K}'}X_i$
where $\mathcal{K}'$ is any randomly selected $(m-1)$-element subset
of $\{1,\ldots,n\}\setminus \{j:X_j=X\}$. With these definitions, we
have $\Pr \bigl[ \mathsf{SINR}^{\mathrm{P1}}\geq 1\bigr] =\Pr
\bigl[\frac{X}{1/\rho +Y}\geq 1\bigr]$. 
For the Rayleigh fading case, in which the link strengths are
i.i.d.\ $\func{Exp}(1)$ random variables, the cumulative
distribution function (cdf) of $X$ (largest of $n$ i.i.d.\
$\func{Exp}(1)$ random variables) can be written explicitly as $
F_{X}(x)=(1-e^{-x})^{n}$. The asymptotic properties of $X$ are well
studied in literature (see \cite{VTL:02} and \cite{SH:05}). For
$s_0=\log n -\log\log n$, it can be shown that $\Pr[X\leq s_0]\to 0$
\cite[eq.~(A4)]{SH:05}. With the help of this property, we proceed
to lower-bound $\Pr \bigl[ \mathsf{SINR}^{\mathrm{P1}}\geq 1\bigr]$
by introducing a real variable $s$ ($0<s\leq s_0$). By the law of
total probability, we have
\begin{align}
\Pr \biggl[\frac{X}{1/\rho +Y}\geq 1\biggr]
           & =\Pr [X> s]\cdot \Pr \biggl[\frac{X}{1/\rho +Y}\geq 1\Bigl|X> s \biggr]  \notag \\
                & \quad +\Pr [X\leq s]\cdot \Pr \biggl[\frac{X}{1/\rho +Y}\geq 1\Bigl|X\leq s\biggr]  \notag \\
           & \geq \Pr [X> s]\cdot \Pr \biggl[\frac{X}{1/\rho +Y}\geq 1\Bigl|X> s\biggr]  \notag \\
           & \geq \Pr [X> s]\cdot \Pr \biggl[\frac{s}{1/\rho +Y}\geq 1\Bigl|X> s\biggr]  \notag \\
           & = \Pr [X> s]\cdot \Pr \biggl[\frac{s}{1/\rho +Y}\geq 1\biggr]  \label{eq:X_Y_indep} \\
           & =\bigl(1-F_{X}(s)\bigr)F_{Y}(s-1/\rho ),  \label{eq:lower_bound_PrSm}
\end{align}
where \eqref{eq:X_Y_indep} follows from the fact that, with $0<s\leq
s_0$ and $\Pr[X\leq s_0]\to 0$, we have $\Pr[X>s]\to 1$, and thus
$\Pr \bigl[\frac{s}{1/\rho +Y}\geq 1\bigl|X> s\bigr]\to \Pr
\bigl[\frac{s}{1/\rho +Y}\geq 1\bigr]$ (which can be trivially shown
by the law of total probability). Note that the lower bound
\eqref{eq:lower_bound_PrSm} suggests a suboptimal scheduling scheme
according to which, each relay schedules the transmission of the
``strongest'' source only if the source's power gain exceeds a
prescribed threshold $s$. The probability of such event is given by
$1-F_X(s)$, and $F_Y (s-1/\rho)$ is a lower bound on the probability
of a successful communication with the relay at a rate of $1$
bit/s/Hz.

The characterization of distribution of the interference term $Y$ in
\eqref{eq:lower_bound_PrSm} needs more care. This is due to the fact
that, conditioned on not being the maximum among $n$ channel
strengths, each interference term in $Y$ is no longer exponentially
distributed, and the interference terms are not independent in
general. However, as shown in Appendix~\ref{app:interference}, these
properties hold asymptotically with $n$. Numerical results in
Appendix~\ref{app:interference} show that these asymptotic trends
are achieved for relatively small values of $n$. Thus, we can
approximate $Y$ as chi-square random variable with $2(m-1)$ degrees
of freedom, whose cdf is thus given by
\cite[eq.~(2.1--114)]{Proakis:4ed}
\begin{equation}
F_{Y}(y)=1-e^{-y}\sum_{k=0}^{m-2}\frac{1}{k!}\,y^{k}.
\label{eq:chi_2m_minus_2}
\end{equation}

Substituting \eqref{eq:lower_bound_PrSm} and
\eqref{eq:chi_2m_minus_2} into \eqref{eq:lower_bound_R_1} yields the
following lower bound on the throughput of Phase $1$.

\smallskip
\begin{lemma}
\label{lem:fix_m_large_n_ul} For any $\rho$, $m$ and $0<s\leq \log
n-\log\log n$, the achievable throughput of the opportunistic relay
scheme in Phase $1$\ is lower-bounded by
\begin{equation}
\label{eq:llower_bound_R_1}
\begin{split}
R_{1}&\geq m\tfrac{n(n-1)\cdots
(n-m+1)}{n^{m}}\bigl(1-(1-e^{-s})^{n}\bigr) \\
& \quad \times \Biggl(1-e^{-(s-{1}/{\rho})}\sum_{k=0}^{m-2}\frac{1}{k!}(s-\tfrac{1}{\rho
})^k\Biggr),
\end{split}
\end{equation}
as $n\to \infty$.
\end{lemma}
\smallskip

A tighter lower bound can be found by maximizing
\eqref{eq:lower_bound_PrSm} over $s$, but we find that little
insight can be gained from this exercise. The tightness of the lower
bound \eqref{eq:llower_bound_R_1} is substantiated by numerical
results shown in Fig.~\ref{fig:lb_R1} of Section~\ref{sec:simulation}.

\smallskip
\begin{remark}
\label{rmk:tradeoff_in_m} Inspecting \eqref{eq:llower_bound_R_1}, we
note that the lower bound on $R_{1}$ exhibits a tradeoff between
quantity and quality of scheduled links. By increasing the number of
relays $m$, one can schedule more simultaneous transmissions, which
is beneficial from the throughput perspective. However, more
transmissions generate more interference, degrading the SINR and
lowering the probability of successful transmissions. In fact, as
shown in Section~\ref{sec:simulation}, not only the lower bound
discussed here, but also the actual throughput $R_{1}$ demonstrates
this tradeoff. The characterization of the best $m$ (in terms of
scaling) that maximizes throughput is pursued in
Section~\ref{sec:large_m_large_n}.
\end{remark}
\smallskip

For the regime of interest, where $n$ is large and $m$ fixed, it can
be trivially shown, e.g., by setting $s=\log n-\log \log n$, that
the above lower bound approaches $m$. Note that this is also the
best we can hope for in Phase $1$, since the $\mathsf{SINR}\geq 1$
constraint to decode a transmitter implies that no more than $m$
sources can be successful.

The following corollary to Lemma~\ref{lem:fix_m_large_n_ul}
follows immediately.

\smallskip
\begin{corollary}
\label{col:Pone_fix_m_large_n} For fixed $m$, $R_{1}\rightarrow m$
as $n\rightarrow \infty $.
\end{corollary}

\subsection{Phase $2$: Relays to Destination Nodes}
\label{sec:Ptwo_fix_m_large_n}

We now develop an exact expression for the sum-rate of the
relay-destination links. This is done by first showing that only a
single relay per destination can produce a required SINR larger than
one, and then computing the probability of the event that a relay is
scheduled and consequently delivers throughput.

\smallskip
\begin{lemma}
\label{lem:fix_m_any_n_dl} For any $\rho_R$, $m$ and $n$, the
achievable throughput of the opportunistic relay scheme at Phase $2$
is given by
\begin{equation}
R_2 = m \left(1- \left( 1-\frac{e^{-1/\rho_R}}{2^{m-1}}\right)^n \right).
\end{equation}
\end{lemma}
\smallskip

Before embarking on the proof, it is worth examining the statistics
of the SINR in \eqref{eq:sinr_dl}. Given the i.i.d. channel model
introduced in the previous section, the SINRs measured at each
destination (cf.~\eqref{eq:sinr_dl}) are of the generic form
$\mathsf{SINR}_{r,j}^{\mathrm{P2}}=\frac{\chi ^{2}(2)
}{1/\rho_{R}+\chi ^{2}(2m-2)}$. With the help of
\eqref{eq:chi_2m_minus_2}, the pdf of the SINR can be shown as
\cite{SH:05}:
\begin{align}
f(x)& =\int_{0}^{\infty }f(x|y)f_{Y}(y)\mathrm{d}y  \notag \\
    & =\frac{e^{-x/\rho _{R}}}{(1+x)^{m}}\left( \frac{1}{\rho _{R}}(1+x)+m-1\right) .  \label{eq:sinr_pdf}
\end{align}%
The corresponding cdf is
\begin{equation}
F(x)=1-\frac{e^{-x/\rho _{R}}}{(1+x)^{m-1}},\quad x\geq 0.
\label{eq:sinr_cdf}
\end{equation}%
Note that the $\mathsf{SINR}_{r,j}^{\mathrm{P2}}$s are i.i.d.\ over
$j=1,\ldots ,n$ (but are not independent over $r=1,\ldots ,m$).

\begin{IEEEproof}
First, we observe that each destination node $j$ has at most one
$\mathsf{SINR}_{k ,j}^{\mathrm{P2}}\geq 1$ for all relays $1\leq k
\leq m$. To see this, assume $\mathsf{SINR}_{k,j}^{\mathrm{P2}}\geq
1$ for some relay $k$, and consider another index $k^{\prime}\neq
k$. From \eqref{eq:sinr_dl}, we have
\begin{equation*}
\xi _{k,j}\geq 1/\rho _{R}+\sum_{\substack{ 1\leq \ell \leq m  \\
\ell \neq k }}\xi _{\ell ,j},
\end{equation*}%
from which it follows that
\begin{equation*}
\xi _{k,j}>\xi _{k^{\prime },j},\quad \forall \,k^{\prime }\neq k.
\end{equation*}%
Therefore
\begin{equation*}
\mathsf{SINR}_{k^{\prime },j}^{\mathrm{P2}}=\frac{\xi _{k^{\prime },j}}{%
1/\rho _{R}+\sum_{\substack{ 1\leq \ell \leq m  \\ \ell \neq k^{\prime }}}%
\xi _{\ell ,j}}<\frac{\xi _{k^{\prime },j}}{\xi_{k,j}}<1.
\end{equation*}%
Thus, each destination node can have at most one \emph{good} relay
as its sender.

Now the sum-rate for Phase $2$ depends on how many relays are
scheduled by destinations. The probability that a relay finds no
destination satisfying $\mathsf{SINR}\geq 1$ is%
\footnote{%
It may seem logical to turn off a relay for which the highest
SINR is still less than one, but we still allow such relays to transmit
(say, control information). This is because, as shown by numerical
results, the performance is limited by the source-relay link.}
\begin{gather}
\Prob{\text{relay $r$ does not receive feedback}}  \notag \\
    \begin{split}
    &=\Pr \left[\mathsf{SINR}_{r,j}^{\mathrm{P2}}\leq 1, \,\, \forall \,j\right] \notag \\
    &=\bigl(F(1)\bigr)^{n}  \notag \\
    &=\left( 1-\frac{e^{-1/\rho _{R}}}{2^{m-1}}\right)^{n}.
    \end{split}
\end{gather}

The throughput of the relay-destination links is given by summing
the probabilities of the relays engaged in transmission. Accounting
for the $1$ bit/s/Hz rate per relay, we have that the average
throughput of the second hop is given by
\begin{align}
R_{2}& =\sum_{r=1}^{m}\Pr [\text{relay $r$ transmits data to a destination}]\cdot 1  \notag \\
     & =m\Bigl(1-\bigl(F(1)\bigr)^{n}\Bigr)  \label{eq:R2_abstract} \\
     & =m\left( 1-\left( 1-\frac{e^{-1/\rho _{R}}}{2^{m-1}}\right) ^{n}\right).\label{eq:R2}
\end{align}
This completes the proof of the lemma.
\end{IEEEproof}

The following corollary ensues by direct computation:

\smallskip
\begin{corollary}
\label{col:Ptwo_fix_m_large_n} For fixed $m$, $R_{2}\rightarrow m$
as $n\rightarrow \infty $.
\end{corollary}
\smallskip

\begin{remark}
\label{rmk:connection} It is interesting at this point to draw a
connection between the scheduling of Phase $2$\ of the opportunistic
scheme proposed here with the random beamforming scheme due to
Sharif and Hassibi in the context of multiple-input multiple-output
broadcast channels (MIMO-BCs) \cite {SH:05}. Seemingly unrelated,
the SINRs of both setups turn out to have the same distribution
(cf.~\eqref{eq:sinr_dl}). To explain this subtlety, note that in the
random beamforming scheme of \cite{SH:05}, a random unitary matrix
$\pmb{\varPhi}$ is applied to the data streams before sending them
over the channel $\pmb{H}$ (hence the terminology \textquotedblleft
random beamforming\textquotedblright ). With the assumption of
i.i.d.\ Rayleigh fading, entries of $\pmb{H}$ follow i.i.d.\
circularly symmetric complex Gaussian random variables
$\mathcal{CN}(0,1)$. By the \emph{isotropic} property of the i.i.d.\
complex Gaussian random matrix $\pmb{H}$, $\pmb{\varPhi H}$ has the
same distribution as $\pmb{H}$ \cite{TV:05}. It follows that the
channel statistics of the random beamforming scheme in the beam
domain are the same as in the original antenna domain. In other
words, the SINR in the beam domain is still of the generic form
\begin{equation}
\mathsf{SINR}=\frac{\chi ^{2}(2)}{1/\rho _{R}+\chi ^{2}(2m-2)},
\label{eq:random_beamforming_SINR}
\end{equation}%
which is the same as in our Phase $2$\ (cf.~\eqref{eq:sinr_dl}).

Despite the mathematical equivalence, our proposed scheme for Phase
$2$\ simplifies the random beamforming scheme in several respects:

\begin{itemize}
\item Random beamforming requires cooperation among the
transmitters to form a beam. Opportunistic relaying operates in a completely
decentralized fashion.

\item Random beamforming requires the feedback of an integer (the beam
index) as well as a real number (the instantaneous SINR). The
proposed opportunistic relaying scheme requires the feedback of only
an index number. This simplification is justified
by~\cite[Th.~2]{SH:05}, which implies that when the system operates
in the limit as $n\rightarrow \infty$ with $m=\Theta(\log n)$, the
aggregate interference from concurrent transmissions eventually
hardens the instantaneous SINR near the value 1. Thus, there is no
longer a need to feed back the SINR value. Furthermore, in terms of
the throughput scaling law (as discussed later in
Section~\ref{sec:large_m_large_n}), this simplification incurs no
loss.
\end{itemize}
\end{remark}

\subsection{Feedback Overhead Analysis}

A detailed study of feedback overhead in the regime of large $n$ and
fixed $m$ is omitted here for the sake of brevity. The calculation
can follow the same steps as in Section~\ref{sec:feedback_overhead},
where we present a detailed analysis of the feedback overhead in the
limiting regime of large $n$ and $m$.

\subsection{Two-Hop Communication}

With the help of Corollaries~\ref{col:Pone_fix_m_large_n} and
\ref{col:Ptwo_fix_m_large_n}, and by taking into account the $1/2$
penalty due to the two hops, the overall system throughput, defined
as $\frac{1}{2}\min \{R_{1},R_{2}\}$, can be readily shown to be
given as follows.

\smallskip
\begin{theorem}
\label{thm:thput_fix_m_large_n} For fixed $m$, the two-hop
opportunistic relaying scheme achieves a system throughput of $m/2$
bits/s/Hz as $n\rightarrow \infty $.
\end{theorem}
\smallskip

Since the proposed scheme works in a decentralized fashion and with
low rate CSI feedback, it is natural to expect some throughput
degradation compared to more intensive schemes. We will show that
the opportunistic relaying scheme exhibits the pre-loglog factor of
the scaling law of the throughput of more intensive schemes. To see
this, we find an information-theoretic upper bound on the achievable
scaling law for the aggregate throughput of \emph{any} two-hop
relaying scheme.

\smallskip
\begin{lemma}
\label{lem:upper_bound} For \emph{any} two-hop relaying
architecture, with fixed $m$ and SNR, the sum rate capacity scales
at most as $\frac{1}{2}m\log \log n$ as $n\to \infty$.
\end{lemma}
\smallskip

\begin{IEEEproof}
In two-hop relay schemes, all data traffic passes through relays.
Therefore, the \emph{best} scheme would be one in which all $m$
relay nodes can cooperate \emph{and} the relays have full CSI (i.e.,
backward as well as forward channel realizations). In such case, the
two-hop communication can be interpreted as MIMO multiple access
channels (MACs) followed by a MIMO-BC. The capacity region of the
MIMO-BC, and the optimality of dirty-paper-coding (DPC) in achieving
the capacity region have been shown in \cite{WSS:06}. Furthermore,
the capacity scaling of the DPC scheme is shown in \cite{SH:05} to
be $m\log \log n$, which is also the capacity scaling for MIMO-MAC
due to the MAC--BC duality \cite{JVG:04}. Now,
Lemma~\ref{lem:upper_bound} follows by taking the two-hop penalty
$1/2$ into account.
\end{IEEEproof}

\begin{remark}
Contrasting Theorem~\ref{thm:thput_fix_m_large_n} to
Lemma~\ref{lem:upper_bound} reveals two different facets of
multiuser diversity. Fundamentally, multiuser diversity gain is a
power gain, e.g., in the Rayleigh fading case, multiuser diversity
schedules the best user for transmission, and boosts the average
power by a factor of $\log n$ \cite{VTL:02}. With the assumption of
relay cooperation, as in Lemma~\ref{lem:upper_bound}, a spatial
multiplexing gain equal to the number of relays $m$ can be readily
achieved (e.g., even by a suboptimal zero-forcing receiver
\cite{TV:05}). Then, multiuser diversity can further boost the rate
of each parallel channel by $\log \log n$, as shown by
Lemma~\ref{lem:upper_bound}. In contrast, with the proposed
opportunistic scheme, where relays operate independently, there is
no guarantee of achieving the multiple parallel channels. Here,
multiuser diversity is used as a mechanism that compensates for the
interference plus noise so that the scheduled link can support $1$
bit/s/Hz. Ultimately, one achieves the linear scaling in $m$. Note
that only with multiuser diversity gain does the SINR of each
\emph{noncooperative} link have the chance to meet the
threshold.\footnote{If one schedules the transmission randomly, the
average receiver SINR can be shown to be $\frac{1}{m-1}$.}
\end{remark}

\smallskip
\begin{remark}
\label{rmk:1bps} At this point, it is worthwhile to revisit the
assumption of $1$ bit/s/Hz fixed transmission rate. According to the
scheduling scheme, receivers select their transmitting nodes by
feeding back their indices. Accordingly, the nodes transmit
independently at $1$ bit/s/Hz. Receivers decode their scheduled
transmitters independently, by treating concurrent interference as
noise. In other words, it is assumed that 1) the transmitters do not
adapt their transmission rate to the instantaneous channel
realizations; and 2) the receivers do not attempt to perform any
interference cancelation. It is reasonable to expect that a higher
throughput can be achieved if we allow rate adaptation and
interference cancelation at the cost of more feedback overhead and
higher computational complexity. However, what
Lemma~\ref{lem:upper_bound} tells us is that the return is at most
the multiplicative factor $\log\log n$. Simulation results in
Section~\ref{sec:simulation} indicate a rather fast convergence to
the asymptotic limits with increasing number of nodes. In a
practical system with finite (but maybe large) $n$, the term $\log
\log n$ is a small number. On the other hand, given the
decentralized scheduling policy adopted here, it is not
straightforward to determine the adaptive transmission rate for each
transmitting node (cf. Remark~\ref{rmk:P1_vs_P2}).
\end{remark}

\section{How Fast Can $m$ Grow?}
\label{sec:large_m_large_n}

As discussed in Remark~\ref{rmk:tradeoff_in_m}, in Phase $1$ there
is a tradeoff between the number of relays $m$ that serve as
conduits between the source and destination nodes and the mutual
interference caused by the transmissions. The same is true for Phase
$2$. This brings up the question: \emph{What is the optimal $m$ that
maximizes the throughput?} This is equivalent to asking the maximum
throughput of the network. In this section, we show that both hops
of the proposed decentralized scheme succeed in achieving
$\Theta(\log n)$ throughput scaling (without taking the feedback
overhead into account). We then quantify the feedback overhead and
conclude that, under the condition that the product of the block
duration and the system bandwidth scales faster than $\log n\log\log n$, the
feedback overhead is negligible and therefore the {\em useful}
throughput of the proposed scheme is given by $\Theta (\log n)$. As
a by-product in characterizing the throughput upper bound of the
first hop, we also conclude that $\Theta(\log n)$ is indeed the best
throughput scaling even if centralized scheduling is allowed. Thus,
as far as throughput scaling is concerned, operating the network in
a decentralized fashion, with local CSI at the receivers and
low-rate feedback to the transmitters, incurs no loss.

\subsection{Phase $1$}
\label{sec:Pone_large_n_large_m}

Earlier, in Section~\ref{sec:Pone_large_n_fix_m}, the lower bound
\eqref{eq:llower_bound_R_1} of the system throughput of Phase $1$
was found. This lower bound was adequate for the discussion in that
section which assumed a large $n$ and fixed $m$. However, in seeking
to determine how the throughput scales with $m$, the lower bound
\eqref{eq:llower_bound_R_1} might considerably underestimate the
true throughput. In light of this, in order to address the question
of optimal $m$, we reason as follows: First, we consider a
genie-aided scheme by relaxing the assumptions of decentralized
relay scheduling. Thus, the throughput scaling for a genie-aided
network with $m$ relays serves as an upper bound on the proposed
decentralized scheme. We show that the throughput scaling law of
this genie-aided scheme is $\Theta (\log n)$. Next, we show that the
lower bound \eqref{eq:llower_bound_R_1} of the proposed
decentralized scheme also achieves the $\Theta (\log n)$ scaling.
Thus, we are able to conclude that the throughput scaling of the
original scheme of Phase $1$\ is given by $\Theta (\log n)$, and is
optimal in a scaling law sense.

\smallskip
\subsubsection{Phase $1$: Upper bound due to genie-aided scheme}

In this subsubsection, we establish the upper bound on the throughput
scaling of Phase $1$ based on the following genie-aided network. The
genie-aided network has access to the full CSI of the network, and
can coordinate the operation of the entire network, i.e.,
centralized scheduling is allowed. Therefore, the genie network can
always achieve the maximum throughput in that, by assumption, it can
enumerate all possible combinations of source-to-relay
transmissions. Nevertheless, we still assume that independent
encoding at the source nodes and independent decoding at the relay
nodes. These constraints are needed to keep the genie-aided
upper-bound result not too loose with respect to the proposed
decentralized network (cf.~Section~\ref{sec:sys_model}). Note that
given a set of channel realizations, the successful source--relay
pairs, in the proposed decentralized scheme, must also be successful
in the genie-aided scheduling scheme. Thus, the throughput of the
genie-aided scheduling scheme upper-bounds the proposed
decentralized scheme.




\smallskip
\begin{theorem}
\label{thm:large_m_Pone} Under the assumption of independent
encoding at the source nodes and independent decoding at the relay
nodes, one cannot achieve
$\frac{\log n}{\log 2}+2$ throughput with probability approaching one.
Conversely, with probability
approaching one, $(1-\epsilon )\frac{\log n}{2\log 2} +2$ throughput
is achievable for all $\epsilon \in (0,1)$.
\end{theorem}
\smallskip

\begin{IEEEproof}[Outline of proof]
The upper bound result is established by showing that, given
$m=\frac{\log n}{\log 2}+2$ relays, with probability approaching
$1$, one cannot find $\frac{\log n}{\log 2}+2$ sources whose
concurrent transmissions to the relays are all successful, even by
enumerating all possibilities of choosing sources and mapping
sources to relays. The achievability result follows from the fact that,
by exhaustive search, with probability of
one, one can find successful concurrent transmissions from $(1-\epsilon
)\frac{\log n}{2\log 2} +2$ sources to $m=(1-\epsilon )\frac{\log
n}{2\log 2} +2$ relays.

See Appendix~\ref{app:proof_thm_2} for detailed proof.
\end{IEEEproof}

\begin{remark}
These results may be of interest in their own right, since, given
the assumptions of independent encoding at the transmitters and
independent decoding at the receivers (reasonable assumptions in ad
hoc networks in which global CSI is not available to enable cooperative
encoding and/or decoding), Theorem~\ref{thm:large_m_Pone}
establishes the upper bound and the achievable throughput scaling
that are valid even if centralized scheduling is allowed. The
$\Theta(\log n)$  throughput exemplifies the interference-limited
nature of the network, and this sub-linear throughput scaling
(compared to other works, e.g., \cite{DH:06,MB:07,OLT:07,CJ:08})
precisely demonstrates the price one has to pay for not having
global CSI knowledge to mitigate the interference. Indeed, recent
works show that if one allows either cooperative decoding between
receivers (see, e.g., \cite{OLT:07}) or cooperative encoding between
transmitters (see, e.g., \cite{CJ:08}), one can indeed avoid/cancel
interference, enabling linear throughput scaling.
\end{remark}

\smallskip
\subsubsection{Achievable throughput scaling of Phase $1$}

Theorem \ref{thm:large_m_Pone} states that, with high probability,
there exists a valid group with $m=(1-\epsilon )\frac{\log n}{2\log
2}+2$ sources such that all transmissions are successful. However,
the proof is nonconstructive: it does not afford insight into how to
find such a set in practice. The proof assumes that there is a genie
with global channel information that can enumerate all possibilities
and select a good one for scheduling. In contrast to the genie-aided
scheme, the opportunistic relaying scheme seeks to operate in a
decentralized manner, and it is not clear whether this operational
simplification incurs a loss in the scaling order of the throughput.
Serendipitously, it can be shown that the ${\log n}$ scaling  is
also met by the lower bound in \eqref{eq:llower_bound_R_1}. To see
this, we examine the asymptotic behavior of
\eqref{eq:llower_bound_R_1}.

Consider the exemplary case of $m=\log n$ and $s=\log n-\log \log
n$. With $n\rightarrow \infty$, the term $\tfrac{n(n-1)\cdots
(n-m+1)}{n^{m}} \rightarrow 1$. The term
$\bigl(1-(1-e^{-s})^{n}\bigr)$ is independent of $m$, and approaches
$1$ for $s=\log n-\log \log n$ as $n\to \infty$. Therefore, a
throughput of $\Theta \left( \log n\right) $ can be achieved as long
as $F_{Y}(s-{1}/{\rho })=\Theta \left( 1\right)$. Indeed, for
$m=\log n$, the interference term $Y$ in (\ref{eq:chi_2m_minus_2}),
by the central limit theorem, can be approximated as Gaussian random
variable with mean and variance both equal to $\log n$. Now, we have
\begin{equation}
F_{Y}(\log n-\log \log n-1/\rho )\approx F_{Y}(\log n)=\frac{1}{2},
\label{eq:R1_asymptotic_lb}
\end{equation}%
due to the symmetry of the Gaussian distribution. Consequently
$R_{1}\approx \frac{1}{2}\log n$. This result implies that for
$m=\log n$ relays, each running the two-hop opportunistic relaying
protocol, it is possible to schedule up to $\log n$ source nodes to
transmit simultaneously, but half of them will fail to satisfy the
SINR requirement due to the multiple access interference. In terms
of throughput, this example yields $\frac{1}{2}\log n$, which
confirms that the scheme is in fact order-optimal in achieving a
throughput of $\Theta \left( \log n\right) $ at Phase $1$.


\subsection{Phase $2$}
\label{sec:Ptwo_large_n_large_m}

In this subsection, we will show that the optimal value of $m$ in
Phase $2$ exhibits a sharp phase transition phenomenon. That is,
$m=\frac{\log n-\log \log n-1/\rho _{R}}{\log 2}+1$ succeeds in
retaining the linearity of $R_{2}$ in $m$, but $m=\frac{\log n+\log
\log n-1/\rho _{R}}{\log 2}+1$ does not. As far as the scaling law
is concerned, this implies that the throughput of Phase $2$ scales
as $\Theta (\log n)$.

\smallskip
\begin{theorem}
\label{thm:Ptwo_large_m} For Phase $2$ of the two-hop opportunistic
relaying scheme, if the number of relays $m=\frac{\log n-\log \log
n-1/\rho_R}{\log 2}+1$, then $R_{2}=\Theta \left( m\right) =\Theta
\left( \log n\right) $. Conversely, if $m=\frac{\log n+\log \log
n-1/\rho_R}{\log 2}+1$, then $R_{2}=o(m)$.
\end{theorem}
\smallskip

\begin{IEEEproof}
For convenience, we repeat $R_{2}$ of \eqref{eq:R2_abstract} and
\eqref{eq:R2}:
\begin{align}
R_{2}& =m\Bigl(1-\bigl(F(1)\bigr)^{n}\Bigr) \\
     & =m\left( 1-\left( 1-\frac{e^{-1/\rho_{R}}}{2^{m-1}}\right) ^{n}\right).\label{eq:R2_reprt}
\end{align}%
With $m=\frac{\log n-\log \log n-1/\rho _{R}}{\log 2}+1$, we have
\begin{equation*}
1-F(1)=\frac{e^{-1/\rho_{R}}}{2^{m-1}}=e^{-(m-1)\log
2-1/\rho_{R}}=\frac{\log n}{n}.
\end{equation*}%
Then,
\begin{align}
\bigl(F(1)\bigr)^{n}& =\left( 1-\frac{\log n}{n}\right) ^{n}=e^{n\log (1-%
                            \frac{\log n}{n})}  \notag \\
                    & =e^{-\log n+O(\frac{\log ^{2}n}{n})}=e^{-\log n+o(\log n)}  \notag \\
                    & =O\left( \frac{1}{n}\right),
\end{align}%
where we have used the fact that, for small $x$, $\log
(1-x)=-x+O(x^{2})$ and $e^{x}=1+O(x)$. Thus, most of the
transmissions meet the SINR threshold (with probability $1-O(1/n)$),
and consequently the throughput $R_{2}$ is given by
$m(1-O(\frac{1}{n}))$. $R_{2}=\Theta (m)=\Theta (\log n)$ follows
readily.

Similarly, when $m=\frac{\log n+\log \log n-1/\rho _{R}}{\log 2}+1$,
we have $1-F(1)=\frac{1}{n\log n}$ and
\begin{align}
\bigl(F(1)\bigr)^{n}& =e^{-\frac{1}{\log n}+O(\frac{1}{n\log
^{2}n})}
\notag \\
& =e^{-\frac{1}{\log n}+o(1/\log n)}  \notag \\
& =1-O(1/\log n).
\end{align}%
Now, in contrast to the case of $m=\frac{\log n-\log \log
n-1/\rho_{R}}{\log 2}+1$, when we increase $m$ to $\frac{\log n+\log
\log n-1/\rho_{R}}{\log 2}+1$, Phase 2 of the two-hop scheme cannot
support a throughput that scales with $m$. With probability one, the
SINRs cannot meet the threshold. In this case, the throughput does
not scale linearly with $m$ anymore, i.e., $R_{2}=o(m)$.
\end{IEEEproof}

This $\Theta (\log n)$ scaling result is consistent with the random
beamforming scheme of \cite{SH:05}, an outcome that is not surprising in
light of the connection discussed in Remark~\ref{rmk:connection}.

\subsection{Feedback Overhead Analysis}
\label{sec:feedback_overhead}

One of the contributions of the paper is the proposal of a two-hop
scheme that alleviates the assumption of full CSI at the
transmitters and the assumption of centralized scheduling. In the
proposed scheme, only CSI at the receivers is employed, but low-rate
feedback from the receivers to the transmitters is assumed to enable
scheduling. In this subsection, we quantify the overhead due to
feedback, and formalize a sufficient condition for which the
overhead is negligible.

\smallskip
\subsubsection{Overhead per fading block}

\smallskip
\smallskip
\noindent{\em \newline Feedback overhead in the first hop:} In the
first hop, each relay schedules one source. Since a total of $n$
sources need to be identified, the feedback overhead per relay is
$\log_2 n$ bits. The overall feedback overhead of all relays is thus
given by $m\cdot \log_2 n$, which scales as $\Theta\bigl((\log
n)^2\bigr)$ since $m=\Theta(\log n)$.

\smallskip
\noindent{\em Feedback overhead in the second hop:} In the second
hop, any destination feeds back the index of a relay only if there
is one relay meeting $\mathsf{SINR}\geq 1$; otherwise, no feedback
is sent. One needs $\Theta(\log_2 m)=\Theta(\log\log n)$ bits to
identify a relay. The number of users that capture a good SINR, and
consequently feed back follows the binomial distribution
$\mathrm{Bi}(n,q)$ with $q$ being the probability that the
destination will provide a feedback. Then, the average overall
feedback overhead is given by the average number of destinations
that feed back, $nq$, times the number of bits of each feedback.

To calculate $q$, we have,
\begin{align}
q&=\Pr[\text{the destination feeds back index}] \notag \\
&=\Pr[\cup_{r=1}^{m}\{\text{the $r$th relay has
$\mathsf{SINR}\geq 1$}\}]\notag \\
&\leq \sum_{r=1}^{m}\Pr[\text{the $r$th
relay has $\mathsf{SINR}\geq 1$}]\notag\\
&=m(1-F(1)) \notag \\
&=\frac{m}{2^{m-1}}e^{-1/\rho}, \label{eq:F_1}
\end{align}
where the last equality is due to \eqref{eq:sinr_cdf}. The quantity
in \eqref{eq:F_1} is of the order of $\Theta\bigl((\log
n)^2/n\bigr)$ when $m=\frac{\log n-\log \log n-1/\rho_R}{\log 2}+1$
(cf.~Theorem~\ref{thm:Ptwo_large_m}).

Finally, the average feedback overhead of the second hop is $O(nq
\log\log n)=O\bigl((\log n)^2 \log\log n\bigr)$.

\begin{figure*}
\centering
\includegraphics[width=\myfigwidth]{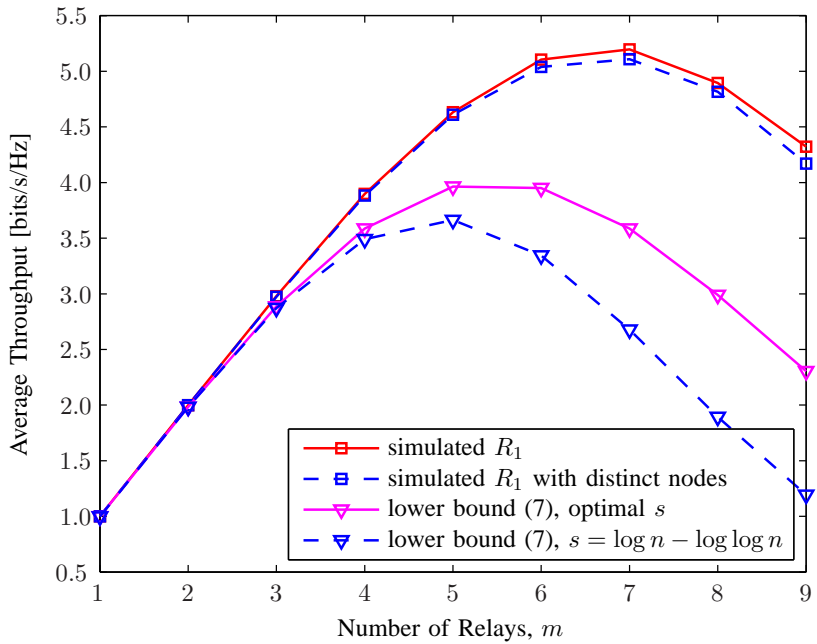}
\caption{\label{fig:lb_R1}First hop average throughput $R_1$ as a function
of the number of relays for $n = 1200$ S--D pairs. From the top:
simulation results utilizing all source node assignments, simulated
results with distinct scheduled nodes, lower bound with optimized
threshold $s$, and lower bound with threshold $s = \log n - \log
\log n$.}
\end{figure*}

\begin{figure*}
\centering
\includegraphics[width=\myfigwidth]{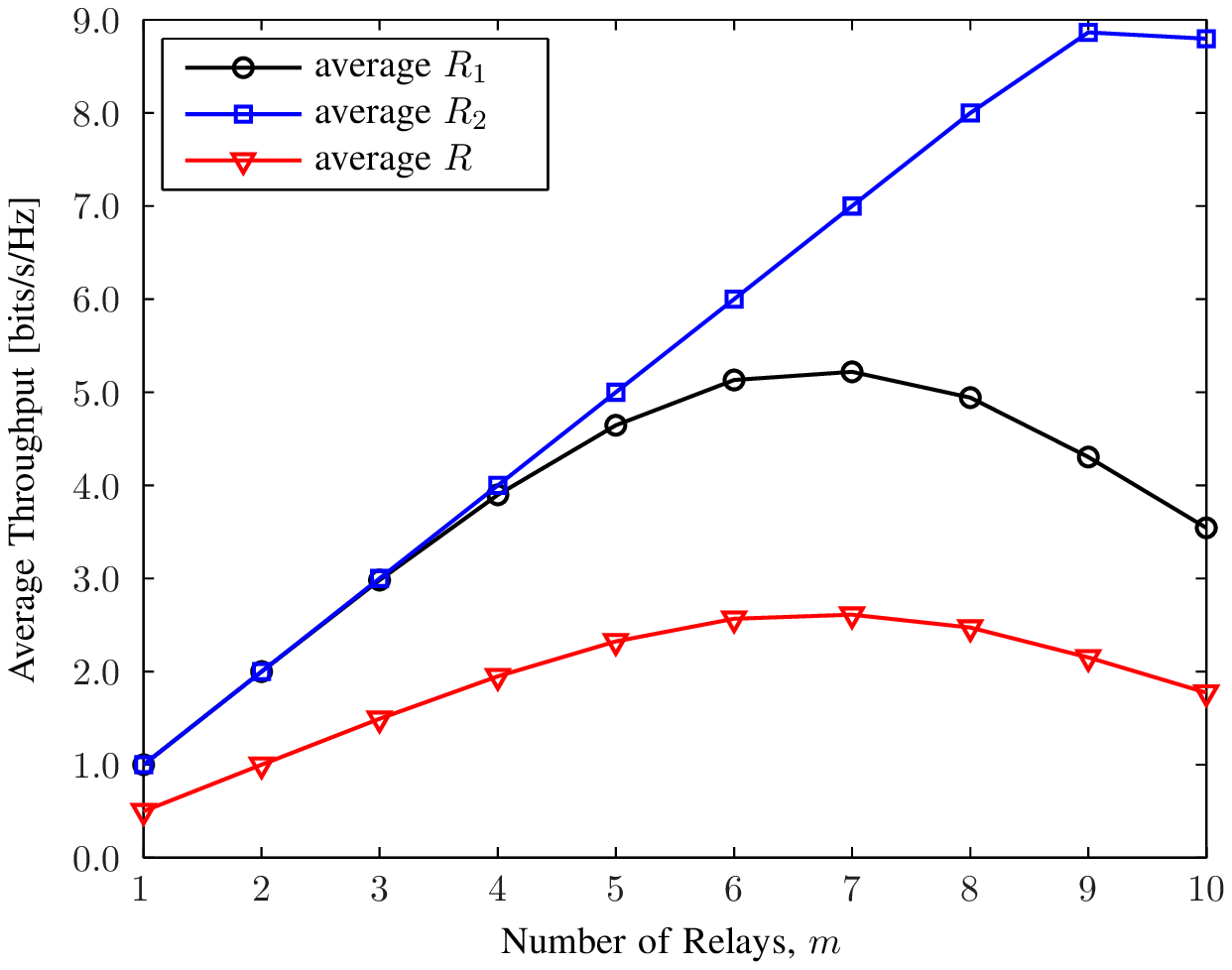}
\caption{\label{fig:thput_n_1200}First hop average throughput $R_1$, second
hop average throughput $R_2$, and average system throughput $R$ as a function of
the number of relays $m$ for $n = 1200$ S--D pairs.}
\end{figure*}

\smallskip
\subsubsection{Condition for $\Theta(\log n)$ useful throughput}
\label{sec:condition}

In this work, a quasi-static fading model is assumed. Specifically,
it is assumed that channel gains are fixed during the transmission
of each hop, and take on independent values at different hops. The
feedback overhead {\em per block} is analyzed above. The throughput
of the system scales with the duration of the
block as well as system bandwidth. That is, the total throughput scales
as $\Theta\bigl(TW\cdot \log n\bigr)$.

Now, we can find conditions of $TW$ for which the feedback overhead
does not imperil the throughput in the sense of throughput scaling.
This is true when the feedback overhead is less than (in terms of
scaling) the total throughput of each hop. Let us look at the second
hop, which has larger feedback overhead. To have $\Theta(\log n)$
useful throughput, we need
$$
(\log n)^2 \log\log n = O(TW\cdot \log n),
$$
which holds whenever $TW=\Omega(\log n \log\log n)$.

In practical system design, $T$ can be as large as the coherence
time of the channel $T_c$, and $W$ can be as large as the coherence
bandwidth of the channel $W_c$. For a typical wireless channel, this
condition is easy to meet. This is due to the fact
that typical wireless channels are \emph{underspread}, that is, they
satisfy $T_cW_c\gg 1$. In typical urban environments, the coherence
bandwidth is of the order of several MHz, and the coherence
time is of the order of milliseconds \cite{TV:05}. Thus, the product
of coherence bandwidth and coherence time is of the order of $10^3$.
As a concrete example, suppose the carrier frequency is $f_c=900$
MHz, and the delay spread is $T_d=1$ $\mu$s. Based on the
definitions of coherence bandwidth and coherence time in
\cite{TV:05}, the coherence bandwidth is given by $W_c=1/(2T_d)=0.5$
MHz. The coherence time depends on velocity $v$, where let us assume
$v=3$ km/h. This leads to a maximum Doppler spread of
$D_s=f_cv/c=2.5$ Hz, and accordingly, to a coherence time of
$T_c=1/4D_s=100$ ms. In this example, $T_cW_c=5\times 10^4$, which
makes $T_cW_c\geq \log n \log\log n$ hold even for extremely large
$n$. For example, for $n=1.0\times 10^8$, $\log n \log\log n=53.7$.

\subsection{Two-Hop Communications}

With the results in previous subsections, we can conclude the
achievable throughput scaling of the scheme in the following
theorem.

\smallskip
\begin{theorem}
\label{thm:max_thput_our_scheme} Under the setup of Section~\ref%
{sec:sys_model}, and given $TW=\Omega(\log n\log\log n)$, the
proposed two-hop opportunistic relaying scheme yields a maximum
achievable throughput of $\Theta \left( \log n\right) $.
\end{theorem}
\smallskip

\begin{remark}
The throughput scaling results in this paper afford a multiuser
diversity interpretation. To see this, it is useful to take a closer
look at the first hop. The power of the signal of each scheduled
link is given by $\log n$ \cite{VTL:02} (due to multiuser
diversity). In the regime where $m$ is fixed
(cf.~Section~\ref{sec:fix_m_large_n}), the signal power can mitigate
the interference power (which is of the order of one), and therefore
each scheduled transmission is successful.
This translates to $m$ bits/s/Hz total throughput of the first hop,
as shown in Corollary~\ref{col:Pone_fix_m_large_n}. In the limiting
operating regime with $m=\Theta(\log n)$ relays, the aggregate
interference for each scheduled link is of the order of $\Theta(\log
n)$. Now, the network saturates as the interference power is of the
same order as the signal power. The system throughput is calculated
as $\Theta(\log n)\cdot \Theta(1)=\Theta(\log n)$, since one has
$\Theta(\log n)$ concurrent transmissions, and each of them has
successful probability of $\Theta(1)$. Further increasing $m$
results in a decreased probability of successful
transmission. Referring back to
Remark~\ref{rmk:tradeoff_in_m}, we see that the optimal $m$ (in the
sense of maximizing system throughput) is given by the order of
multiuser diversity. The interpretation of the throughput scaling in
terms of multiuser diversity is discussed in more detail in
\cite{CH:08}.
\end{remark}

\section{Numerical Results}
\label{sec:simulation}

In this section, we provide some numerical examples produced by
simulations of the proposed opportunistic relaying scheme under
Rayleigh fading. Throughout these examples, the SNR for both hops is
set at $10$ dB ($\rho =\rho _{R}=10\text{ dB}$).

We examine in Fig.~\ref{fig:lb_R1} the average throughput $R_{1}$ of
the first hop of the protocol and its various lower bounds. The
figure contains four curves. The two simulation curves were obtained
by averaging throughputs over $2,000$ channel realizations. The
\textquotedblleft simulated $R_{1}$\textquotedblright\ curve was
obtained using all assignments of source nodes, while the curve
marked \textquotedblleft simulated $R_{1}$ with distinct
nodes\textquotedblright\ represents only assignments of distinct
source nodes. The other two lower bounds shown are computed with
\eqref{eq:llower_bound_R_1}: one is obtained by optimizing
\eqref{eq:llower_bound_R_1} over $s$ (numerically); the other lower
bound is for $s=\log n-\log \log n$. Three observations are
noteworthy relative to Fig.~\ref{fig:lb_R1}. First, both the
simulated throughput and the analytical lower bound
\eqref{eq:llower_bound_R_1} exhibit linearity with respect to $m$,
consistent with the analysis of
Section~\ref{sec:Pone_large_n_fix_m}. Second, it is observed that
when $m$ exceeds a certain value (in this case, $6$), the throughput
$R_{1}$ starts to fall off. Noting that $\log 1200=7$, this effect
is consistent with the analysis in
Section~\ref{sec:Pone_large_n_large_m} that established that the
linear increase in throughput with the number of relays holds only
as long as $m$ is of the order $\log n$. Third, the lower bound of
$R_{1}$ of \eqref{eq:llower_bound_R_1} becomes loose when $m$ grows.
The development leading to (\ref{eq:lower_bound_PrSm}) suggests two
possible reasons for this behavior. The first is that the
computation of $\Pr \left[ N_{m}\right]$ is based on only distinct
source nodes. However, the close match between the two simulation
curves in Fig.~\ref{fig:lb_R1} eliminates this possibility. It
follows then that the bound is loosened due to the series of
lower-boundings of $\Pr \bigl[ \frac{X}{1/\rho+Y}\bigr] $ leading to
\eqref{eq:lower_bound_PrSm} being too conservative.

In Fig.~\ref{fig:thput_n_1200}, we illustrate throughputs $R_{1}$
and $R_{2}$, as well as the corresponding system throughput of the
full scheme given by $R=\frac{1}{2}\min \{R_{1},R_{2}\}$. As
discussed in Section~\ref{sec:sys_model}, the transmissions over the
second hop are destined to be successful, since they are scheduled
based on SINR\ measurements at the destination nodes, whereas the
transmissions over the first hop are not guaranteed to be successful
since they are based only on SNR measurements. As a consequence, we
observe from Fig.~\ref{fig:thput_n_1200} that $R_{1}$ is lower than
$R_{2}$, and is the bottleneck to the system throughput, i.e.,
$R=\frac{1}{2}R_{1}$. In addition, we observe that the optimal
number of relays for Phase $2$\ is consistent with the analysis of
Theorem~\ref{thm:Ptwo_large_m} in
Section~\ref{sec:Ptwo_fix_m_large_n}. Nevertheless, both $R_{1}$ and
$R_{2}$ display the linearity in $m$ as predicted by
Corollaries~\ref{col:Pone_fix_m_large_n} and
\ref{col:Ptwo_fix_m_large_n} in Section~\ref{sec:fix_m_large_n}.

\begin{figure*}
\centering
\includegraphics[width=\myfigwidth]{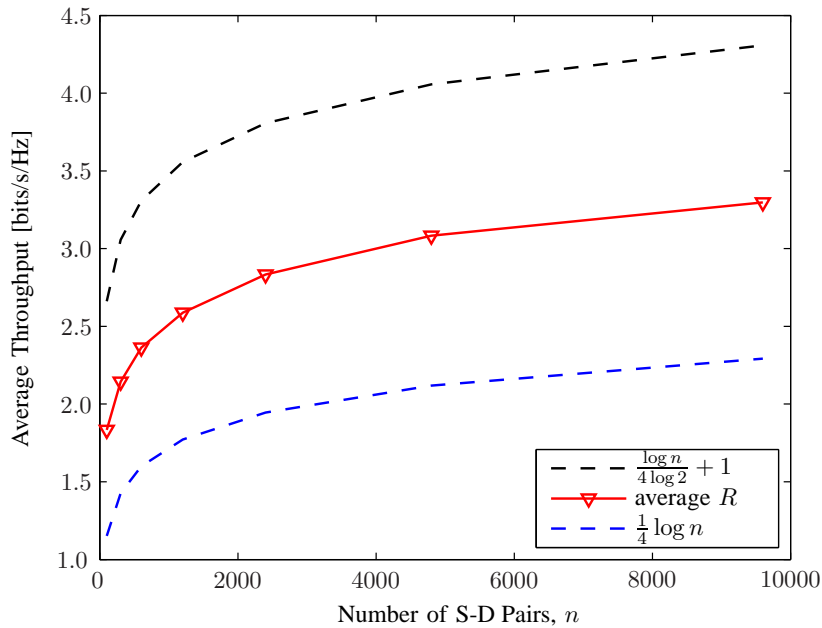}
\caption{\label{fig:optimal_thput}Simulated average system throughput of the
proposed scheme as a function of the number of S--D pairs $n$ and for
optimized number of relays $m$. Also shown are a genie upper bound
and the lower bound $\frac{1}{4} \log n$.}
\end{figure*}

\begin{figure*}
\centering
\includegraphics[width=\myfigwidth]{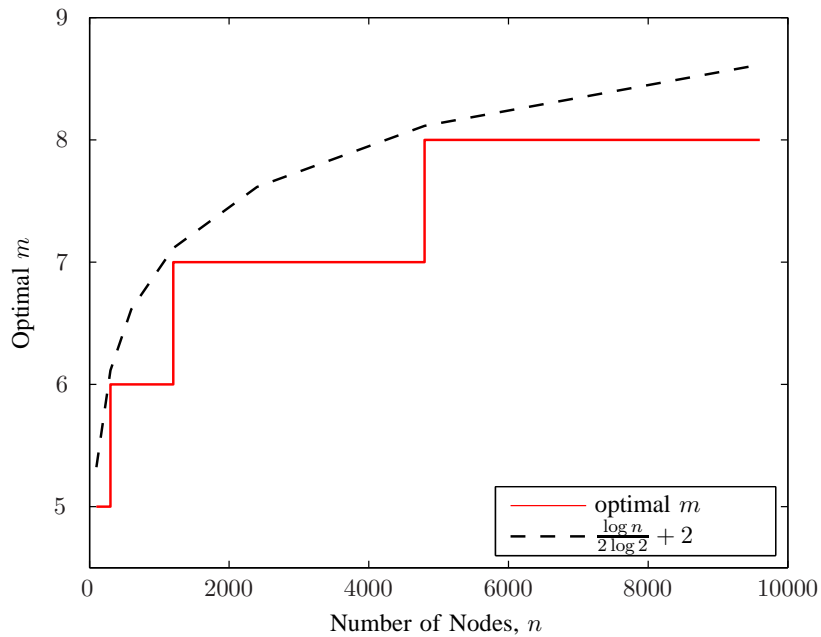}
\caption{\label{fig:optimal_m}Optimal value of $m$ that maximizes
the average throughput, and $\frac{\log n}{2 \log 2} + 2$ curve
(cf.~Theorem~\ref{thm:large_m_Pone})).}
\end{figure*}

The total throughput of the two-hop opportunistic relaying scheme is
shown in Fig.~\ref{fig:optimal_thput} as a function of the number of
nodes $n$. We observe that the throughput exhibits the $\log n$
trend, as predicted by Theorem~\ref{thm:max_thput_our_scheme}. In
fact, the system throughput curve can be perfectly approximated by
$0.36\log n$. Since the system throughput is always limited by Phase
$1$, i.e., $R=\frac{1}{2}\min \{R_{1},R_{2}\}= \frac{1}{2}R_{1}$, we
also plot two bounds of $\frac{1}{2}R_{1}$ for reference. More
specifically, the genie bound $\frac{\log n}{4\log 2}+1$
(cf.~Theorem~\ref{thm:large_m_Pone}) serves as an upper bound, and
the $\frac{1}{4}\log n$ curve from \eqref{eq:R1_asymptotic_lb}
serves as a lower bound for the system throughput. In
Fig.~\ref{fig:optimal_m}, the optimal value of the number of relays,
$m$, is shown versus the number of nodes, $n$.
Comparing the values of $m$ from the curve, with the value
$\frac{\log n}{2\log 2}+2$, which is the bound on the number of
relays for the genie scheme in Theorem~\ref{thm:large_m_Pone}, we
observe that the optimal $m$ is very close to that of the
genie-bound. This explains why the scheme can harness large portions
of throughput as promised by Theorem 4.

\section{Other Considerations: Relay Cooperation and Delay}
\label{sec:discussions}

One of the key contributions of this work is to propose an
opportunistic relaying scheme that features decentralized relay
operations and practical CSI assumptions. In this section, we
discuss the case in which relays are allowed to cooperate in
encoding/decoding. In order to isolate the impact of
the relay cooperation on the two-hop scheme, we leave the
CSI assumptions unchanged. Specifically, it is assumed that the relays have
full CSI knowledge of the source--relay link, but have only partial
index-valued CSI knowledge of the relay--destination link via feedback.
This discussion will help identify the fundamental limits of the
opportunistic relaying scheme. Finally, we briefly address the issue
of network delay.

\subsection{Cooperative Relays}
\label{sec:cooperate_relays}

In the proposed opportunistic relaying scheme, we assume the relays
perform independent decoding (in the first hop) and independent
encoding (at the second hop). In particular, the relays treat the
received interference as noise, and no attempt is made to cancel
the mutual interference caused by concurrent transmissions. As a
consequence, the system is \emph{interference limited}. In this
subsection, we address the question: \emph{How does cooperation
between relays in decoding/encoding change the scheduling operation
and the throughput scaling?} For example, it is conceivable that the
relays could be implemented as infrastructure nodes that are
connected to a wired backbone. This setup has been referred to as a
\emph{hybrid network} (see for example \cite{ZV:05} and references
therein).

\begin{table*}[htb!]
\caption{Dependence of Throughput Scaling on Relay Cooperation in
Encoding/Decoding and CSI Knowledge of the Relay--Destination Link}
\label{tbl:summary_relay_coop}\renewcommand{\arraystretch}{1.3}
\centering
\begin{minipage}{\textwidth}
\centering
\begin{tabular}{c|c|c|c}
\hline
\bfseries Scenario\footnote{Case $1$: independent decoding/encoding at the relays; perfect CSI in the first hop and partial CSI in the second hop\\
\phantom{---..}Case $2$: cooperative decoding/encoding at the relays; perfect CSI in the first hop and partial CSI in the second hop\\
\phantom{---..}Case $3$: cooperative decoding/encoding at the relays; perfect CSI in the first hop and perfect CSI in the second hop} & \bfseries Throughput Scaling of $R_1$ & \bfseries %
Throughput Scaling of $R_2$ & \bfseries Throughput Scaling of $R$ \\
\hline\hline Case $1$ & $\Theta (\log n)$ & $\Theta
(\log n)$ & $\Theta (\log n)$ \\
Case $2$ & $\Theta (n)$ & $\Theta (\log n)$ & $%
\Theta (\log n)$ \\
Case $3$ & $\Theta (n)$ & $\Theta (n)$ & $\Theta (n)$ \\
\hline
\end{tabular}%
\end{minipage}
\end{table*}

When the relays are allowed to fully cooperate in decoding/encoding,
they can be considered to be a multi-antenna array. Accordingly, the
first and second hops are equivalent to a MIMO MAC with receiver
CSI, and a MIMO BC with partial transmitter CSI, respectively. Now,
the scheduling in Phase $1$ can be simplified. It is well-known that
for the MIMO MAC, the sum-capacity can be achieved by allowing all
users to transmit. The receiver can retrieve the data via some
sophisticated signal processing algorithm, e.g., MMSE-SIC
(minimum-mean square estimator with successive interference
cancelation) \cite{VG:97}. The optimal scaling in the large $n$ and
fixed $m$ regime is given by $m\log \log n$. However, if we seek to
achieve only linear scaling in $m$, it suffices to schedule
\emph{any} $m$ source nodes for transmission. With high probability,
the resulting $m\times m$ channel is well-conditioned, and a spatial
multiplexing gain of $m$ is achieved \cite{TV:05}. In contrast,
Phase $2$ does not benefit from the cooperation of relays. This is
because, with only partial transmitter CSI, and since destination
nodes are not allowed to collaborate, arbitrarily selecting $m$
destination nodes cannot yield a throughput linear in $m$
\cite{SH:05}.

The impact of relay cooperation on throughput scaling exhibits
similar behavior to that demonstrated for scheduling. Phase $1$\
benefits from relay cooperation, and in principle $R_{1}=\Theta (n)$
is possible (note that the capacity of a $n\times n$ MIMO channel
scales linearly with $n $ \cite{telatar:99}); Phase $2 $\ is still
bounded by $\Theta (\log n)$, and becomes the bottleneck of the
two-hop scheme. The reason for this is that, due to the lack of full
CSI at the relays, there is no way to generate more than $\Theta
(\log n)$ parallel channels. The reader is referred to \cite{SH:05}
for a discussion of the impact of CSI knowledge on MIMO downlink
channels.

We summarize the discussion of relay cooperation in
Table~\ref{tbl:summary_relay_coop}. In the first two scenarios, we
examine the impact of cooperation in decoding/encoding at the relays
on system throughput by fixing the CSI assumptions to perfect
receiver CSI in the first hop and partial CSI in the second hop.
Specifically, case $1$ corresponds to the setup considered in
Section~\ref{sec:sys_model}, where the relays perform independent
decoding (in the first hop) and independent encoding (in the second
hop), and case $2$ allows for cooperation among relays in
decoding/encoding. In the comparison, we also include the optimistic
scenario, case $3$, where the relays are assumed to have full CSI
knowledge of both the source--relay link, and the relay--destination
link and the relays are allowed to cooperate. In this case,
$\Theta(n)$ throughput is obtained in both hops, a result not surprising from
the MIMO theory \cite{telatar:99,Foschini:96}. From the table, one
can readily identify that CSI plays a critical role in determining
if linear throughput scaling is achievable. This observation
justifies our study on throughput scaling based on the seemingly
pessimistic, yet practical, assumptions on CSI knowledge in this
paper.

It is important to point out that in the above discussion, the focus
is on CSI. In cases where perfect CSI is
available at the relays, but cooperative decoding/encoding is not
available (e.g., due to nodes located randomly), different
conclusions can be drawn. For example, one can operate the two-hop
amplify-and-forward scheme \cite{DH:06} to achieve $\Theta(n^{1/2})$
throughput scaling. A detailed discussion of the case of perfect CSI
but no cooperation between relays is outside the scope of this
paper. It is also important to point out that the discussion applies
only to the underlying Rayleigh fading model. For other fading
models, the opportunistic relaying scheme may exhibit a different
scaling law. See \cite{CHSPS:08} for discussions of
throughput scaling under more general fading models.

\subsection{Delay Considerations}

There is always a tension between opportunistic scheduling and delay
considerations \cite{VTL:02}. The delay issue is more salient in the
two-hop scheme than in the cellular setup \cite{VTL:02}, because
packets transmitted by one particular source in Phase $1$\ must be
buffered at a relay, until that relay schedules the original
destination during Phase $2$. While one can partially relieve the
problem by, say, prioritizing the destination in cases when a relay
receives multiple requests from multiple destinations (including the
destination of interest, of course), the delay may still be large.
The detailed study of end-to-end delay is currently underway.

\begin{figure*}[!t]
\centering
\includegraphics{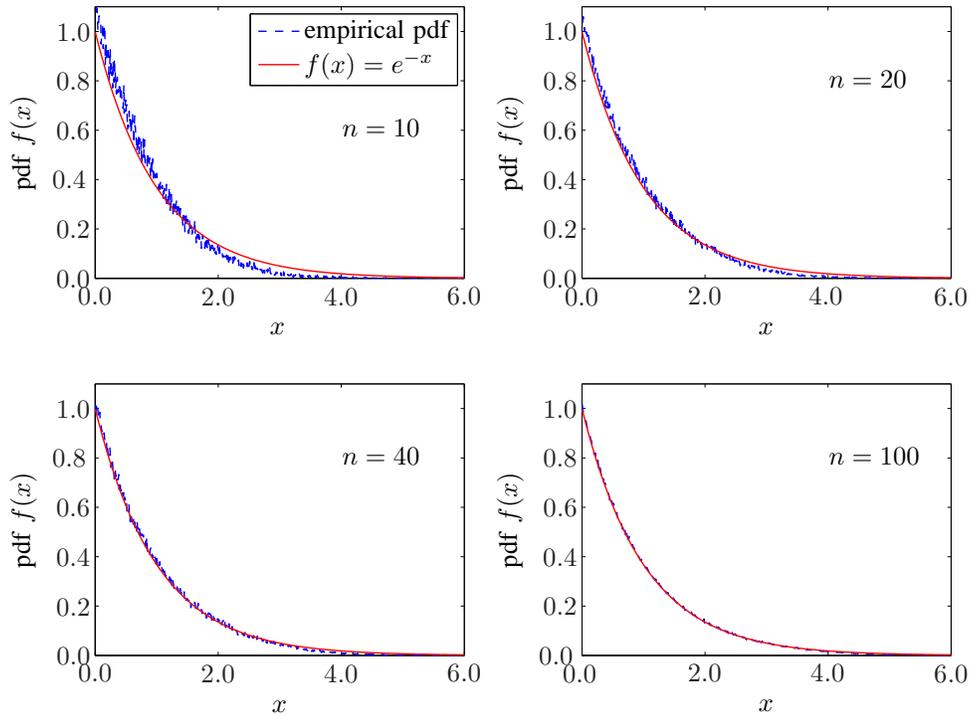}
\caption{Empirical pdf $f(x_i\vert \text{$X_i$ is not the maximum})$
with $n=10,20,40$ and $100$ respectively and the pdf of a standard
exponential random variable, $f(x)=e^{-x}, \; x\geq 0$.}
\label{fig:asyp_Exp}
\end{figure*}

\section{Conclusion}
\label{sec:conclusion}

In this work, we have proposed an opportunistic relaying scheme that
alleviates the demanding assumptions of central scheduling and CSI
at transmitters. The scheme entails a two-hop communication
protocol, in which sources can communicate with destinations only
through half-duplex relays. The key idea is to schedule at each hop
only a subset of nodes that can benefit from multiuser diversity. To
select the source and destination nodes for each hop, relays operate
independently with receiver CSI only, and with an index-valued
feedback to the transmitter. The system throughput has been
characterized for the operating regime in which $n$ is large and $m$
is relatively small. In this case, the proposed scheme achieves a
system throughput of $m/2$ bits/s/Hz, while the upper bound with
full cooperation among relays and full CSI is $(m/2)\log \log n$.
Moreover, we have further shown that, given that the product of the
block duration and the system bandwidth scales as $\Omega(\log n
\log\log n)$, the achievable throughput scaling of the proposed
decentralized scheme is given by $\Theta (\log n)$, which is the
optimal scaling even if centralized scheduling is allowed. Thus,
operating the network in a decentralized fashion, with only CSI at
the receivers and low-rate feedback to the transmitters, incurs no
penalty. Finally, compared to the linear throughput scaling results
reported in the literature (see, e.g., \cite{OLT:07} and \cite{CJ:08})
with more optimistic CSI assumptions, this work quantifies the price
that one has to pay for not being able to mitigate interference. The
delay behavior of the proposed opportunistic relaying scheme is left
for future work.

\appendices

\section{Characterization of Interference $Y$ of \eqref{eq:X_Y_indep}}
\label{app:interference}

In this appendix, we characterize the statistical properties of the
interference term $Y$ in \eqref{eq:X_Y_indep}. More specifically, it
is shown that, asymptotically in $n$, each individual term that
comprises $Y$ has an exponential distribution, and all interferers
are asymptotically independent. It is also illustrated by numerical
results that these asymptotic trends are achieved quickly, enabling
the approximation of $Y$ as a chi-square random variable with
$2(m-1)$ degrees of freedom.

For notional convenience, we denote the channel connections from $n$
sources to the relay as $X_1,\ldots,X_n$. According to the
scheduling of Phase $1$, for each time-slot, i.e., each realization
of $X_1,\ldots,X_n$, the desired signal strength is the maximum
among all connections. The interference term $Y$ is the summation of
$(m-1)$ out of the remaining $(n-1)$ channel connections.

We first show that each interferer is asymptotically exponentially
distributed. By the law of total probability, for events $B$ and
$A$, we have
\begin{equation}
\Pr [B]=\Pr [B|A]\Pr [A]+\Pr [B|\overline{A}]\Pr [\overline{A}],
\end{equation}%
where $\overline{A}$ denotes the complement of the event $A$. Now
define the event $B$ as $\{X_{i}\leq x_{i}\}$, and $A$ as
$\{$$X_{i}$ is not the maximum$\}$. Then we have
for the cdfs%
\begin{equation}
F_{X_i}(x_{i})=F_{X_i}(x_{i}|A)\Pr
[A]+F_{X_i}(x_{i}|\overline{A})\Pr [\overline{A}].
\end{equation}%
In our i.i.d.\ model, by symmetry, each node has probability of
$1/n$ to be the maximum, i.e., $\Pr [A]=1-{1}/{n}$. Thus, the above
equation can be written as
\begin{equation}
F_{X_i}(x_{i})=F_{X_i}(x_{i}|A)\underbrace{(1-\frac{1}{n})}_{\rightarrow 1}+\underbrace{%
F_{X_i}(x_{i}|\overline{A})\frac{1}{n}}_{\rightarrow 0}.
\end{equation}%
Therefore, we have
\begin{equation}
F_{X_i}(x_{i}|A)\rightarrow F_{X_i}(x_{i})\quad \text{as
}n\rightarrow \infty,
\end{equation}%
and thus, asymptotically, each interferer is still exponentially
distributed.

While the above results are of an asymptotic nature, numerical
result shows that they hold for practical values of $n$ as well. For
example, in Fig.~\ref{fig:asyp_Exp}, the empirical pdf $f(x_i|
\text{$X_i$ is not the maximum})$ is plotted together with the pdf
of $\func{Exp}(1)$, i.e., $f(x)=e^{-x}, \; x\geq 0$, for various
values of $n$. It is seen that the empirical pdf is well
approximated by the standard exponential distribution.

Next, we show that the interferers are asymptotically independent.
Define $A$ as $\{$none of $X_{1},\ldots ,X_{m-1}$ is the
maximum$\}$. The event $\overline{A}$ is then $\{$at least one of
$X_{1},\ldots ,X_{m-1}$ is the maximum$\}$. Again, by the law of
total probability,
\begin{equation}
\label{eq:total_prob_all_interfe}
\begin{split}
F(x_{1},\ldots ,x_{m-1})& =F(x_{1},\ldots ,x_{m-1}|A)\Pr [A] \\
& \quad +F(x_{1},\ldots ,x_{m-1}|\overline{A})\Pr [\overline{A}].
\end{split}%
\end{equation}%
Due to the underlying i.i.d.\ assumption, $\Pr
[A]=(1-1/n)\bigl(1-1/(n-1)\bigr)\cdots\bigl(1-1/(n-m+2)\bigr)\rightarrow
1$ and $\Pr [\overline{A}]\rightarrow 0$ when $n$ is large and $m$
small relative to $n$. Then it follows readily from
\eqref{eq:total_prob_all_interfe} that in the regime of interest

\begin{equation}
f(x_{1},\ldots ,x_{m-1}|A)\rightarrow f(x_{1},\ldots
,x_{m-1})=\prod_{i=1}^{m-1}f(x_{i}).
\end{equation}%
Therefore, asymptotically, all interferers are independent.

Combining the facts that, in the regime of interest, 1) each
interferer is exponentially distributed and 2) all interferers are
independent, the aggregate interference $Y$ can thus be modelled as
chi-square with $2(m-1)$ degrees of freedom. Numerical result shows
that this approximation is accurate for values as low as $n=40$.

\section{Proof of Theorem 2}
\label{app:proof_thm_2}

Here, we prove Theorem~\ref{thm:large_m_Pone} of
Section~\ref{sec:Pone_large_n_large_m}. For convenience, the theorem
is repeated below.

\smallskip
\indent {\em Theorem 2:} Under the assumption of independent
encoding at the source nodes and independent decoding at the relay
nodes, one cannot achieve
$\frac{\log n}{\log 2}+2$ throughput with probability approaching one.
Conversely, with probability
approaching one, $(1-\epsilon )\frac{\log n}{2\log 2} +2$ throughput
is achievable for all $\epsilon \in (0,1)$.
\smallskip

\begin{IEEEproof}
The proof relies on the probabilistic method \cite{AS:00}. The basic
idea of the probabilistic method is that in order to prove the
existence of a structure with certain properties, one defines an
appropriate probability space of structures and then shows that the
desired properties hold in this space with positive probability.
This method of proof has been seen in various subjects of
information theory, for instance, see \cite[Ch.~8]{Etkin:06}, which
studies the bandwidth scaling problem in the context of spectrum
sharing. The line of our proof follows \cite{Etkin:06}.

The upper bound is established by the genie-aided scheduling, which
performs an exhaustive search for the maximum concurrent successful
transmissions. Specifically, in testing whether $m$ bits/s/Hz is achievable,
the genie-aided scheme
enumerates all $m$-element subset of source nodes and tests whether the
resulting $m$ transmissions to the $m$ relays are all successful.
According to the genie-aided scheme, we define the probability space
$\Omega =\bigl\{(\mathcal{A},\pi ):\mathcal{A}\subset \{1,\ldots
,n\},\abs{\set{A}}=m,\text{ }\pi \text{ is any permutation on
$\{1,\ldots ,m\}$}\bigr\}$, where $\mathcal{A}$ denotes a random
$m$-set of all $n$ source nodes and $\pi $ denotes any possible
$m$-to-$m$ mappings from $m$ source nodes in $\mathcal{A}$ to $m$
relays. Let $B_{\mathcal{A}}^{\pi}$ be the event $\{$all nodes in
$\mathcal{A}$ can transmit simultaneously and successfully under
mapping rule $\pi $$\}$ and $I_{\mathcal{A}}^{\pi}$ the
corresponding indicator random variable, i.e., $I_{\mathcal{A}}^{\pi
}=\mathbf{1}\Bigl(\frac{\gamma _{i,R_{\pi}(i)}}{1/\rho
+\sum_{\substack{ t\in
\mathcal{A}  \\ t\neq i}}\gamma _{t,R_{\pi}(i)}}\geq 1,\quad \forall i\in \mathcal{%
A}\Bigr)$, where the subscript $R_{\pi}(i)$ denotes the
corresponding relay for source $i$ under mapping rule $\pi$. For any
$\pi$, we have $\{R_{\pi}(i),\forall i\in
\mathcal{A}\}:=\mathcal{R}=\{1,\ldots,m\}$. Finally, define the
number of valid sets that satisfy the SINR threshold as
$X(m)=\sum_{\mathcal{A}}\sum_{\pi }I_{\mathcal{A}}^{\pi }$.


Then
\begin{align}
\mathbb{E}[I_{\mathcal{A}}^{\pi }]& =\Pr [B_{\mathcal{A}}^{\pi }]  \notag \\
& =\Pr \Bigl[\mathsf{SINR}_{i,R_{\pi}(i)}^{\mathrm{P1}}\geq 1,\quad
\forall i\in
\mathcal{A}\Bigr]  \notag \\
& =\biggl(\Pr \Bigl[\mathsf{SINR}_{i,R_{\pi}(i)}^{\mathrm{P1}}\geq 1\Bigr]\biggr)%
^{m}  \label{eq:sinr_iid_over_r} \\
& =(p_{m})^{m},  \label{eq:p_m}
\end{align}%
where \eqref{eq:sinr_iid_over_r} follows from the fact that for $i,
j \in \mathcal{A}, i\neq j$,
$\mathsf{SINR}_{i,R_{\pi}(i)}^{\mathrm{P1}}$ and
$\mathsf{SINR}_{j,R_{\pi}(j)}^{\mathrm{P1}}$ are i.i.d. The term
$p_{m}=1-F(1)$ in \eqref{eq:p_m} is the probability that a
transmission is successful when there are $m$ concurrent
transmissions, and $F(\cdot )$ is the cdf of the SINR computed in
\eqref{eq:sinr_cdf}.

The linearity of the expectation yields
\begin{equation}
\mathbb{E}\left[ X(m)\right] =\binom{n}{m}m!\,(p_{m})^{m}.
\label{eq:expt_Xm}
\end{equation}

Then, the upper bound is established by showing $\Pr [X(m)\geq
1]\rightarrow 0$ when $m=\frac{\log n}{\log 2}+2$. This can be seen
from Markov's inequality:
\begin{align}
\Pr [X(m)\geq 1]& \leq \mathbb{E}[X(m)]  \notag \\
& =\frac{n!}{(n-m)!}\,(p_{m})^{m}  \notag \\
& \leq (np_{m})^{m}\leq (\tfrac{ne^{-1/\rho }}{2^{m-1}})^{m}  \notag \\
& =e^{m(\log n-(m-1)\log 2-1/\rho )}  \notag \\
& \leq e^{m(\log n-(m-1)\log 2)}.  \label{eq:to_follow1}
\end{align}

Now substituting $m=\tfrac{\log n}{\log 2}+2$ into
\eqref{eq:to_follow1}, we have
\begin{align}
\Pr [X(m)\geq 1]& \leq e^{-\log n+o(\log n)}  \notag \\
& =O\bigl(\tfrac{1}{n}\bigr).  \label{eq:cannot_find_set}
\end{align}%
What \eqref{eq:cannot_find_set} tells us is that when $m=\tfrac{\log
n}{\log 2}+2$, the probability of finding a set of $m$ nodes for
concurrent successful transmissions decreases to zero as $n$
increases. Since the transmission rate is fixed at $1$ bit/s/Hz, it
is equivalent to concluding that, with probability approaching one,
$\tfrac{\log n}{\log 2}+2$ bits/s/Hz throughput is not achievable.

\begin{figure*}[!b]
\vspace*{4pt}
\hrulefill
\normalsize
\setcounter{MYtempeqncnt}{\value{equation}}
\setcounter{equation}{33}
\begin{align}
\mathbb{E}\Bigl[\Bigl.I_{\mathcal{A}}I_{\mathcal{A}^{\prime
}}\Bigr|\substack{\set{A}\in \Omega'\\ \set{A}'\in \Omega'\\
\abs{\set{A}\cap \set{A}'}=q}\Bigr]& =\mathbb{E}\Biggl[\prod_{k\in
\mathcal{A}}\mathbf{1}\biggl(\frac{\gamma _{k,R(k)}}{\sigma
^{2}/P+\sum
_{\substack{ t\in \mathcal{A}  \\ t\neq k}}\gamma _{t,R(k)}}\geq 1\biggr)%
\prod_{\ell \in \mathcal{A}^{\prime }}\mathbf{1}\biggl(\frac{\gamma _{\ell
,R(\ell )}}{\sigma ^{2}/P+\sum_{\substack{ t\in \mathcal{A}%
^{\prime }  \\ t\neq \ell }}\gamma _{t,R(\ell )}}\geq 1\biggr)%
\Biggr]  \notag \\
& \leq \mathbb{E}\Biggl[\prod_{k\in \mathcal{A}}\mathbf{1}\biggl(\frac{%
\gamma _{k,R(k)}}{\sigma ^{2}/P+\sum_{\substack{ t\in \mathcal{A}  \\ t\neq
k }}\gamma _{t,R(k)}}\geq 1\biggr)\prod_{\ell \in \mathcal{A}^{\prime
}\setminus \mathcal{A}}\mathbf{1}\biggl(\frac{\gamma _{\ell ,R(\ell )}}{\sigma ^{2}/P+\sum_{\substack{ t\in \mathcal{A}^{\prime }  \\ %
t\neq \ell }}\gamma _{t,R(\ell )}}\geq 1\biggr)\Biggr]
\label{eq:etkin1} \\
& \leq \mathbb{E}\Biggl[\prod_{k\in \mathcal{A}}\mathbf{1}\biggl(\frac{%
\gamma _{k,R(k)}}{\sigma ^{2}/P+\sum_{\substack{ t\in \mathcal{A}
\\ t\neq k }}\gamma _{t,R(k)}}\geq 1\biggr)\prod_{\ell \in
\mathcal{A}^{\prime }\setminus
\mathcal{A}}\mathbf{1}\biggl(\frac{\gamma _{\ell ,R(\ell )}}{\sigma
^{2}/P+\sum_{\substack{ t\in \mathcal{A}^{\prime
}\setminus \mathcal{A}  \\ t\neq \ell }}\gamma _{t,R(\ell )}}\geq 1%
\biggr)\Biggr]  \label{eq:etkin2} \\
& =\mathbb{E}\Biggl[\prod_{k\in \mathcal{A}}\mathbf{1}\biggl(\frac{\gamma
_{k,R(k)}}{\sigma ^{2}/P+\sum_{\substack{ t\in \mathcal{A}  \\ t\neq k}}%
\gamma _{t,R(k)}}\geq 1\biggr)\Biggr]\mathbb{E}\Biggl[\prod_{\ell \in
\mathcal{A}^{\prime }\setminus \mathcal{A}}\mathbf{1}\biggl(\frac{\gamma
_{\ell ,R(\ell )}}{\sigma ^{2}/P+\sum_{\substack{ t\in \mathcal{A}%
^{\prime }\setminus \mathcal{A}  \\ t\neq \ell }}\gamma _{t,R(\ell
)}}\geq 1\biggr)\Biggr]  \notag \\
& =(p_{m})^{m}(p_{m-q})^{m-q}  \label{eq:Delta_ub}
\end{align}
\setcounter{equation}{\value{MYtempeqncnt}}
\end{figure*}

Next we look at achievability. In proving the achievability result,
we consider a variant of the genie-aided scheme used above. Here,
the scheme divides the total $n$ sources into $m$ groups
$\mathcal{G}_i, i=1,\ldots,m$ with each group having $n/m$ sources.
Each group is associated with one relay node. For example, as
illustrated in Fig.~\ref{fig:group}, without loss of generality, we
can label the total source nodes from $1$ to $n$ and assign sources
$\{1,\ldots, n/m\}$ to $\mathcal{G}_1$, $\{n/m+1,\ldots, 2n/m\}$ to
$\mathcal{G}_2$, and so on. In testing whether $m$ concurrent
successful transmissions are possible, each relay chooses one source
from its own group.\footnote{Similar scheme has been considered by
Etkin \cite[Ch.~8]{Etkin:06} in the context of characterizing the
bandwidth scaling of spectrum sharing systems. Our setup is
different from Etkin's scheme in that the number of nodes in each
group is a function of $m$, which is not the case in
\cite{Etkin:06}.} Following the scheme, we define the sample space
$\Omega' =\bigl\{\mathcal{A}:\abs{\mathcal{A}}=m, R(i)\neq R(j) \,\,
\forall i,j\in \mathcal{A}\bigr\}$, where $R(i)$ denotes the index
of the relay associated with the group to which source $i$ belongs.
Also define $I_{\mathcal{A}}$ as the indicator random variable of
the event $\{$transmission from source $i$ to relay $R(i)$ is
successful, $\forall \, i\in\mathcal{A}$$\}$. Finally, let
$X'(m)=\sum_{\mathcal{A}\in \Omega'}I_{\mathcal{A}}$.

\begin{figure}[h!]
\centering
\includegraphics{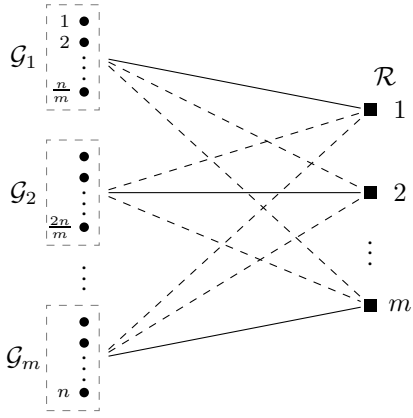}
\caption{Illustration of the genie-aided scheme used in the
achievability proof. Source nodes are divided into $m$ groups
$\mathcal{G}_i, i=1,\ldots,m$, each with $n/m$ nodes. Each group
associates with one relay node and the nodes in the group have
common receiver (i.e., the associated relay). $\mathcal{A}$ is
formed by selecting one node from each group.} \label{fig:group}
\end{figure}

To prove the achievability, we seek to find a lower bound on
$\Pr[X'(m)\geq 1]$, or equivalently, an upper bound on
$\Pr[X'(m)=0]$. We need the following probabilistic tool from
\cite{Etkin:06}.

\smallskip
\begin{lemma}
\label{lem:extended_Jason_inequality} Let $\mu=\mathbb{E}[X'(m)]$
and $\Delta =\sum_{\mathcal{A}\in \Omega'}\sum_{\substack{
\mathcal{A'}\in \Omega'  \\ \mathcal{A}\cap \mathcal{A'}\neq 0}}
\mathbb{E}[I_{\mathcal{A}}I_{\mathcal{A}^{\prime }}]$. Then,
\begin{equation}
\Pr [X(m)=0]\leq e^{-\frac{\mu ^{2}}{\Delta }}.
\end{equation}
\end{lemma}
\smallskip

\begin{IEEEproof}
Following the explanation in the proof of \cite[Th.~10]{Etkin:06},
the proof of the lemma follows in a fairly straightforward way from
\cite[Lemma~7]{Etkin:06}. We skip the details for the sake of
saving space.
\end{IEEEproof}

The proof of achievability is more involved than that of the upper
bound. This is because $I_{\mathcal{A}}$'s are generally not
independent. In the upper bound proof, however, dependence among
$I_{\mathcal{A}}$'s is irrelevant due to the linearity of the
expectation. The quantity $\Delta$ in
Lemma~\ref{lem:extended_Jason_inequality} is a measure of the
pairwise dependence between the $I_{\mathcal{A}}$'s. Note that, in
the case when $I_{\mathcal{A}}$'s are all independent, the lemma
reduces to $\Pr[X'(m)=0]=e^{-\mu}$, a result which can be reached by
direct probability calculations.

We begin with $\mu$:
\begin{align}
\mu &= \mathbb{E}[X'(m)] =\mathbb{E}\bigl[{\textstyle \sum_{\mathcal{A}\in\Omega'} } I_{\mathcal{A}}\bigr] \notag \\
    &={\textstyle \sum_{\mathcal{A}\in\Omega'} } \mathbb{E} [I_{\mathcal{A}}] \label{eq:linear_of_E}\\
    &= \Bigl(\frac{n}{m}\Bigr)^m (p_m)^m, \label{eq:A_iid}
\end{align}
where \eqref{eq:linear_of_E} follows from linearity of expectation,
and \eqref{eq:A_iid} is due to the fact that, for any
$\mathcal{A}\in \Omega'$, all relays see i.i.d. channel
realizations. The term $p_m=\frac{e^{-1/\rho }}{2^{m-1}}$
(cf.~\eqref{eq:sinr_cdf}) denotes the probability of successful
decoding when there are $m$ concurrent transmissions in total.

To compute $\Delta$, let us start with computing the expectation of
$I_{\mathcal{A}}I_{\mathcal{A}^{\prime }}$ conditioned on
$\{\set{A}\in \Omega', \set{A}'\in \Omega', \abs{\set{A}\cap
\set{A}'}=q\}$. In the expressions (shown at the bottom of the page),
\eqref{eq:etkin2} upper-bounds \eqref{eq:etkin1} by neglecting
the interference coming from the sources belonging to
$\mathcal{A}^{\prime }\cap \mathcal{A}$ in the second term of the product. In so
doing, the two products $\prod_{k\in \mathcal{A}}\mathbf{1}(\cdot )$
and $\prod_{\ell \in \mathcal{A}^{\prime }\setminus
\mathcal{A}}\mathbf{1}(\cdot )$ in \eqref{eq:etkin2} involve
independent random variables now and therefore are independent (note
that this is not true in \eqref{eq:etkin1}). A minimal example is
illustrated in Fig.~\ref{fig:Delta_example}. The upper-bounding in
\eqref{eq:etkin2} can be thought of as reducing the number of
concurrent transmissions from $m$ to $m-q$ by keeping the elements
$\{t:t\in \mathcal{A}^{\prime }\cap \mathcal{A}\}$ silent. The
probability of successful transmission when there are $m-q$
concurrent transmissions, denoted as $p_{m-q}$, can be shown to be
$p_{m-q}=\frac{e^{-1/\rho }}{2^{m-q-1}}$.
\addtocounter{equation}{3}

\begin{figure}[h!]
\centering
\includegraphics{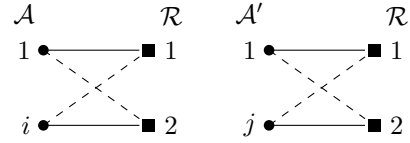}
\caption{Example of $\mathcal{A}=\{1,i\}$ and
$\mathcal{A}^{\prime}=\{1,j\}$, $i\neq j$. We can upper-bound the
SINR of source $j$ in $\mathcal{A}^{\prime}$ as
$\frac{\protect\gamma_{j,2}}{1/\protect\rho+\protect\gamma_{1,2}}\leq
\frac{\protect\gamma_{j,2}}{1/\protect\rho}$, which now is
independent of the SINRs of source nodes in $\mathcal{A}$.}
\label{fig:Delta_example}
\end{figure}

Now, we proceed with $\Delta$. In particular, we have
\begin{align*}
\Delta &= \sum_{\mathcal{A}\in \Omega'}\sum_{\substack{
\mathcal{A'}\in \Omega'  \\ \mathcal{A}\cap \mathcal{A'}\neq 0}}
\mathbb{E}[I_{\mathcal{A}}I_{\mathcal{A}^{\prime }}] \\
&= \sum_{q=1}^{m}\sum_{\set{A}\in \Omega'}\sum_{\substack{ \set{A}'\in \Omega'  \\ %
\abs{\set{A}\cap \set{A}'}=q}}\mathbb{E }\Bigl[%
\Bigl. I_{\mathcal{A}}I_{\mathcal{A}^{\prime}}\Bigr|
\substack{\set{A}\in \Omega'\\ \set{A}'\in \Omega'\\
\abs{\set{A}\cap \set{A}'}=q} \Bigr] \\
&= \sum_{q=1}^{m}\Bigl(\frac{n}{m}\Bigr)^m \binom{m}{q}
\Bigl(\frac{n}{m}-1\Bigr)^{m-q} (p_m)^{m}(p_{m-q})^{m-q}.
\end{align*}

In order to apply Lemma~\ref{lem:extended_Jason_inequality}, we
check $\frac{\Delta}{\mu^2}$:
\begin{equation*}
\frac{\Delta}{\mu^2} = \sum_{q=1}^{m}\frac{\binom{m}{q}
\bigl(\frac{n}{m}-1\bigr)^{m-q} }{
\bigl(\frac{n}{m}\bigr)^m}\frac{(p_{m-q})^{m-q}}{(p_m)^m} =
\sum_{q=1}^{m} a_q,
\end{equation*}
where $a_q=\frac{\binom{m}{q} (\frac{n}{m}-1)^{m-q} }{
(\frac{n}{m})^m}\frac{(p_{m-q})^{m-q}}{(p_m)^m}$. Now on defining
$b_q=a_{q+1}/a_q$, we have that
$b_q=\frac{(m-q)e^{1/\rho}}{(\frac{n}{m}-1)(q+1)}2^{2(m-q-1)}$
decreases with $q$, $b_q \leq b_1$. Therefore, $a_q \leq
b_1^{q-1}a_1$.

On setting $m=(1-\epsilon)\frac{\log n}{2\log 2}+2$ for any $\epsilon \in (0,1)$,
we have
\begin{align*}
b_1 &= \frac{(m-1)e^{1/\rho}}{2(\frac{n}{m}-1)} 2^{2(m-2)} \\
&= e^{\log (m-1) - \log (\frac{n}{m}-1) +2(m-2)\log 2+1/\rho-\log 2 } \\
&= e^{-\epsilon \log n +o(\log n)} \\
&= O\Bigl(\frac{1}{n^\epsilon}\Bigr).
\end{align*}

Furthermore, with $m=(1-\epsilon)\frac{\log n}{2\log 2}+2$, we have
\begin{align*}
\frac{\Delta}{\mu^2} &= \sum_{q=1}^{m} a_q \leq a_1 \sum_{q=1}^{m} b_1^{q-1}
\leq \frac{a_1}{1-b_1} \\
&= \frac{\binom{m}{1}(\frac{n}{m}-1)^{m-1}}{(\frac{n}{m})^{m}}\frac{(p_{m-1})^{m-1}}{%
(p_m)^m}\frac{1}{1-b_1}\\
&< \frac{m^2}{n}
\frac{(p_{m-1})^{m-1}}{%
(p_m)^m}\frac{1}{1-b_1} \\
& =  \frac{m^2}{n} \frac{%
e^{1/\rho}\,2^{2(m-1)}}{1-b_1} \\
&= e^{2\log m -\log n + 2(m-1)\log 2 -\log (1-b_1)+1/\rho} \\
&= e^{-\epsilon \log n +o(\log n)} \\
&= O\Bigl(\frac{1}{n^\epsilon}\Bigr).
\end{align*}

Finally, Lemma~\ref{lem:extended_Jason_inequality} yields
\begin{equation}
\Pr [X(m)=0]<e^{-n^\epsilon}. \label{eq:find_set}
\end{equation}

In words, \eqref{eq:find_set} tells us is that when
$m=(1-\epsilon)\tfrac{\log n}{2\log 2}+2$ for any $\epsilon \in (0,1)$, the
probability of not finding a set of $m$ nodes for concurrent
successful transmissions decreases to zero as $n$ increases. In
other words, the probability of finding $m$ concurrent successful
transmissions with $m=(1-\epsilon)\tfrac{\log n}{2\log 2}+2$
approaches $1$. Again, since the transmission rate is fixed at $1$
bit/s/Hz, it is equivalent to concluding that, with probability
approaching one, $m=(1-\epsilon)\tfrac{\log n}{2\log 2}+2$ bits/s/Hz
throughput is achievable by the genie-aided scheme.

This achievability result, together with the upper bound
\eqref{eq:cannot_find_set}, completes the proof of the theorem.
\end{IEEEproof}

\section*{Acknowledgment}

The authors wish to thank Dr.~Raul Etkin for helpful discussions on
the proof of Theorem~\ref{thm:large_m_Pone}, and for pointing out
\eqref{eq:etkin2} in particular. The authors would also like to thank the Associate Editor
Sennur Ulukus and the anonymous reviewers for their
thoughtful comments and valuable suggestions which helped
us improve the initial submitted version of this manuscript.

\bibliographystyle{IEEEtran}

\end{document}